\documentclass[floats,floatfix,amssymb,prd,superscriptaddress,nofootinbib,twocolumn,aps,eqsecnum]{revtex4}

\usepackage[utf8]{inputenc}

\usepackage{amsmath, amsthm, amsfonts, amssymb, latexsym}
\usepackage{graphicx}
\usepackage{url}
\usepackage{xspace}

 \usepackage[usenames]{color}

\usepackage[unicode]{hyperref}
\hypersetup{
    unicode=true,          
    pdftitle={Resolving the relative influence of strong field spacetime dynamics and MHD on circumbinary disk physics},
    pdfcreator={LaTeX},          
    pdfproducer={LaTeX},         
    pdfkeywords={General Relativity, Accretion disks, MHD, Black Holes},
    colorlinks=false,            
  }

\newcommand{\harm}{{\sc Harm3d}\xspace}   

\newcommand{\dF}{{^{^*}\!\!F}}

\newcommand{\bF}{{\bf F}}
\newcommand{\bU}{{\bf U}}

\newcommand{\del}{{\partial}}
\newcommand{\bsq}{{{||b||^2}}}

\newcommand{\prim}{{{\mathbf{P}}}}

\newcommand{\mhdonepnold}{{MHD1PN\_l}\xspace}
\newcommand{\mhdonepnnew}{{MHD1PN\_H}\xspace}
\newcommand{\mhdtwopn}{{MHD2.5PN}\xspace}

\newcommand{\hydroonepn}[1]{{hydro1PN\_a#1\xspace}}
\newcommand{\hydrotwopn}[1]{{hydro2.5PN\_a#1\xspace}}

\begin{document}

\title{Resolving the relative influence of strong field spacetime dynamics and MHD on circumbinary disk physics}

\author{Miguel~Zilh\~ao}

\affiliation{
  Center for Computational Relativity and Gravitation, 
  School of Mathematical Sciences, and School of Physics and Astronomy,
  Rochester Institute of Technology, 
  Rochester, NY 14623, USA
}

\affiliation{
  Departament de F\'{\i}sica Fonamental \& Institut de Ci\`{e}ncies del Cosmos, 
  Universitat de Barcelona, 
  Mart\'{\i} i Franqu\`{e}s 1, E-08028 Barcelona, Spain
}

\author{Scott~C.~Noble}
\affiliation{
  Center for Computational Relativity and Gravitation, 
  School of Mathematical Sciences, and School of Physics and Astronomy,
  Rochester Institute of Technology, 
  Rochester, NY 14623, USA
}
\affiliation{ Department of Physics \& Engineering Physics, University of Tulsa, Tulsa, OK, 74104}

\author{Manuela~Campanelli}
\author{Yosef~Zlochower}
\affiliation{
  Center for Computational Relativity and Gravitation, 
  School of Mathematical Sciences, and School of Physics and Astronomy,
  Rochester Institute of Technology, 
  Rochester, NY 14623, USA
}


\begin{abstract}
In this paper we evolve magnetized and unmagnetized circumbinary accretion disks around supermassive black hole binaries in the relativistic regime. We use a post-Newtonian expansion to construct an analytical spacetime and determine how the order of the post-Newtonian (PN) expansion affects the dynamics of the gas. We find very small differences in the late-time bulk dynamics of non-magnetized hydrodynamic evolutions between the two spacetimes down to separations of approximately $40GM/c^2$ where $M$ is the total mass of the binary. For smaller separations, the differences due to PN-order become comparable to differences caused by using initial data further from equilibrium. For magnetized gas, MHD stresses, which drives the accretion dynamics, tends to mask all higher order PN effects even at separations of $20GM/c^2$, leading to essentially the same observed electromagnetic luminosity. This implies that our calculations of the EM signal may be robust down to small binary separations. Our investigation is the first to demonstrate how the level of PN accuracy affects a circumbinary disk's evolution and informs us of the range in separation within which to trust the PN approximation for this kind of study. We also address the influence the initial conditions and binary separation have on simulation predictions.
\end{abstract}

\maketitle


\section{Introduction}
\label{sec:intro}

Supermassive black holes (SMBH) in the nuclei of galaxies are
understood to play a key role in the construction of galaxies, as
evidenced by the strong correlations between their masses and their
host galaxies' stellar bulge masses and velocity dispersions
\cite{Magorrian1998, Gebhardt2000, FerrareseMerritt2000,
KormendyHo2013, SilkRees1998, Fabian1999}.

Because today's galaxies are generally thought to have been assembled
from mergers of smaller galaxies, SMBH binaries may be a common
occurrence in the nuclei of the merged galaxies
\cite{Vol03, Vol07, Schnitt07}.  Subsequent to the galactic merger,
dynamical
friction from dark matter and baryonic matter (e.g. stars and/or
gas) should bring the SMBHs close to the center of mass of the merged
galaxy, where a variety of angular momentum loss mechanisms may bring
them still closer together~\cite{BBR80}.  Once the orbital separation
shrinks to $\lesssim 1000 r_g$ (where $r_g \equiv GM/c^2$ and $M$ is
the mass of the binary),
gravitational radiation drives the orbit of the binary which rapidly
inspirals down to merger.

Despite relatively few observations of SMBH mergers to date
~\citep{Deane:2014jqa, Bogdanovic:2014cua}, we know that the rate of these events should 
be at least a few per year. Programs, such as
the {\sl Panoramic Survey Telescope and
Rapid Response System} (Pan-STARRS), which is already in operation,
and the planned {\sl Large Synoptic Survey Telescope} (LSST) will be able
to search for these events using electromagnetic (EM) signals. Similarly,
 pulsar
timing arrays can probe for these events in gravitational waves (GW).
In the long term, EM signatures for SMBH mergers will also help us 
pinpoint GW sources from future space missions such the European 
New GW Observatory (NGO), also known as eLISA
\citep{2013GWN.....6....4A, AmaroSeoane:2012je, Seoane:2013qna},
determine the redshift luminosity distance relationship to large
redshifts and can be used to constrain GW parameter inference.
For all these reasons, it is therefore important to provide 
accurate predictions of the EM emission of SMBH mergers in 
the relativistic regime.

Modeling the gas dynamics near merging SMBHs can be extremely challenging.
Hydrodynamic (HD) and magnetohydrodynamic (MHD) simulations 
of accretion disks around SMBHs binary systems have been carried out in the
Newtonian regime~\citep{Macfadyen:2006jx,Shi:2011us,Noble:2012xz, RoedigSesana14, Farris14a, Shi14, Nixon11, Bankert14} 
when the binary is well-separated and in the late-inspiral and merger phase 
~\citep{Bode10, Bode12, Pal10b, Farris10, Farris12, Farris14a,
Giacomazzo12, Gold:2013zma}. However, until the work
of Noble et al~\cite{Noble:2012xz}, the inspiral regime remained unexplored.

Noble et al~\cite{Noble:2012xz} introduced the idea of using an
analytical spacetime of an inspiralling black-hole binary using
post-Newtonian approximations to solve the field equations
of general relativity~\cite{Blanchet:1998vx}.
This allowed for simulations of disks for more than a hundred orbits using 
the \harm MHD code developed by Noble~\cite{GMT03,Noble:2008tm}, far longer than would be
practical with typical full GRMHD codes, which solve the MHD and
gravitational field equations numerically.
 The \harm code is now a mature code 
that solves the MHD equations on arbitrary dynamical
spacetimes in arbitrary coordinate systems. Here we use 
a spherical grid that is adapted to the geometry of the disk which is ideal
to study circumbinary accretion dynamics.

In the inspiral regime, far from the sources  [$r_g/r = GM/(rc^2) \ll
1$], where the BH motion is  slow
[$(v/c)^2 \ll 1$], the post-Newtonian (PN) approximation gives a very good description of
spacetime dynamics~\cite{Blanchet:1998vx, Will:2011nz}.   The PN approximation is 
an asymptotic series in powers of these small  quantities, characterized by the order 
to which it is taken [e.g., 3PN means up to terms $\sim (r_g/r)^3$].  The PN metric
takes energy loss from the  binary into account, accurately modeling
both the energy loss and inspiral of the binary.

In a previous paper~\citep{Noble:2012xz}, we used a spacetime model that is accurate up 
to 2.5PN order  (i.e., including terms up to $\sim (r_g/r)^{5/2}$)
but describing the binary orbital evolution to 3.5PN.  We showed that circumbinary disks can, in
part, track the inspiral of a SMBH binary even at late stages of its
evolution.  The resulting metric was not valid
very close to the BHs, and consequently, we excised any material that
fell within 1.5 binary separations. 
More recently, we developed a new technique in which the entire
relevant region of spacetime is covered by a number of individual zones, each of
them based on an analytic approximation appropriate to its particular conditions ~\cite{Mundim:2013vca}.
This will allow us to simulate how the gas falls onto the binary, distribute itself
into two mini-disks around each BH,  and evolves within these disks.
This will be the subject of a separate upcoming paper. 

In this paper, our goal is to explore the region of validity of the PN
metric, where higher-order PN corrections become important, and where
the PN spacetime needs to be supplanted by a numerical one.
Our approach is based on using approximations 
where they are appropriate in order to employ the most 
computationally-intensive methods only on the domains in
which they are essential.

Even though our spacetime metric is able to cover the full simulation
domain, in this project, for simplicity, we avoid evolving the gas in
the neighborhood of the BHs, and thus excise a spherical
domain which includes the binary from our calculation.  
We follow the same excision procedure as was used before 
in \cite{Noble:2012xz} and excise the innermost 
1.5 binary separations of the domain.  Indeed, in
order to include the BHs in the computational domain while keeping the
overall problem size at a practical level, 
one would need to introduce a new, spherical-like, nonuniform coordinate 
system~\cite{Zilhao:2013dta}. 

To quantify how the PN order of accuracy affects the
evolutions of non-magnetized and magnetized gas, we perform here a sequence
of inviscid hydrodynamic evolutions of nearly identical disks using
Newtonian, 1PN, and 2.5PN order metrics at separations from
$100 r_g$ to as small as $15 r_g$. 
Here we will examine both the transient behavior of
the disk (which is very sensitive to the PN order), the
quasi-equilibrium state of non-magnetized gas (which is less sensitive),
and the quasi-equilibrium state of magnetized gas (which is even less
sensitive). Ultimately, we find that the PN approximation can be used
to evolve MHD disks down to binary BH (BBH) separations as small as $20 r_g$,
leading to robust calculations of the observed EM luminosity.

In the rest of this paper, we use the  conventions of Misner,
Thorne and Wheeler~\cite{Misner73} for the spacetime metric throughout.
 We use the Greek letters $(\alpha, \beta,
\cdots)$ to denote spacetime indices, and Latin letters $(i, j, \cdots)$
to denote spatial indices. The metric is denoted $g_{\mu \nu}$ and
it has signature $(-,+,+,+)$. We use the geometric
unit system, where $G=c=1$, with the useful conversion factor $1 M_{\odot} =
1.477 \; {\rm{km}} = 4.926 \times 10^{-6} \; {\rm{s}}$.

\section{Simulation details}
\label{sec:techniques}

\subsection{Quasi-Equilibrium Initial Data and Spacetime Treatment}
\label{sec:comparison}

\begin{figure*}[tbp]
  \centering
  \includegraphics[width=0.4\textwidth]{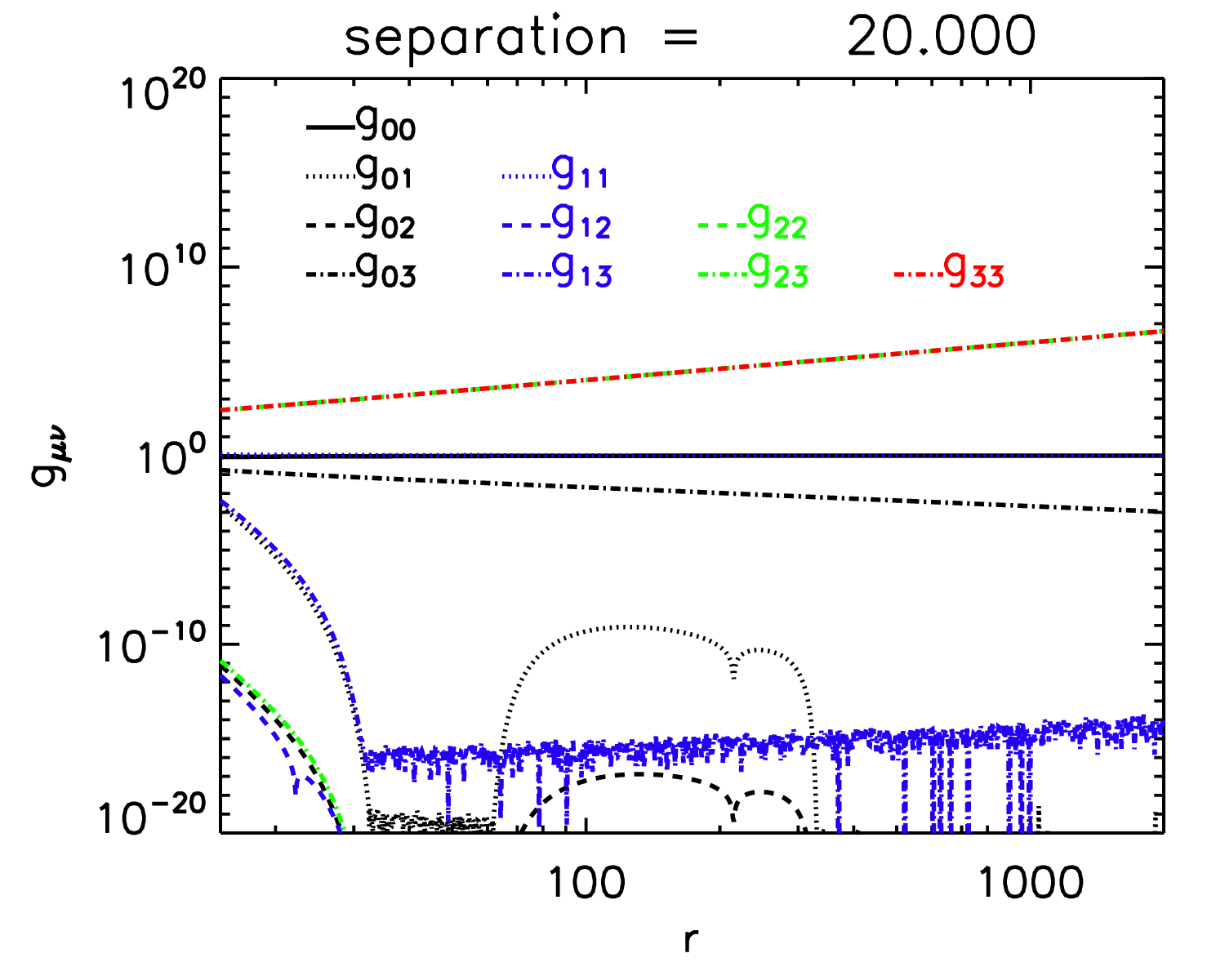} 
  \includegraphics[width=0.4\textwidth]{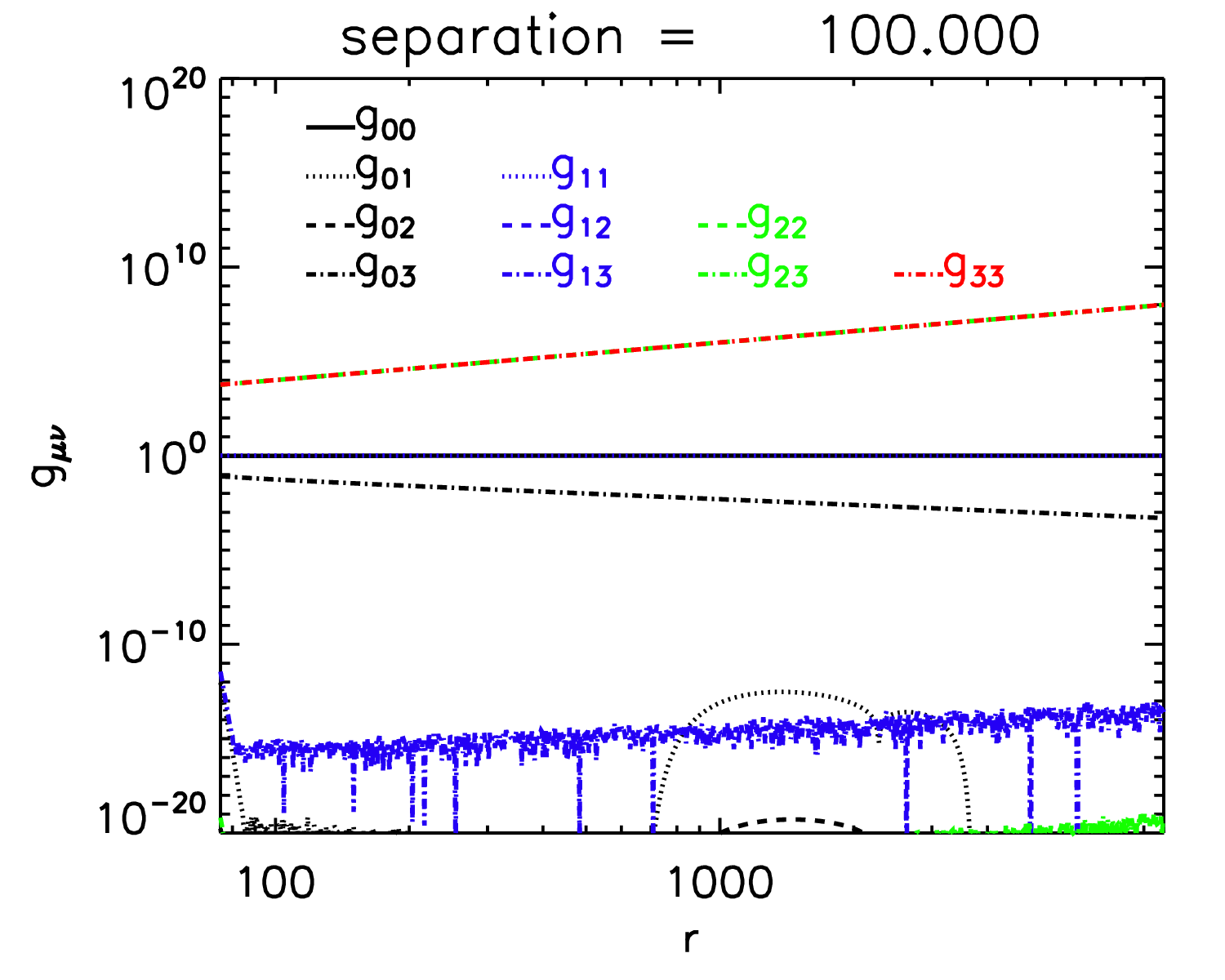}\\
  \includegraphics[width=0.4\textwidth]{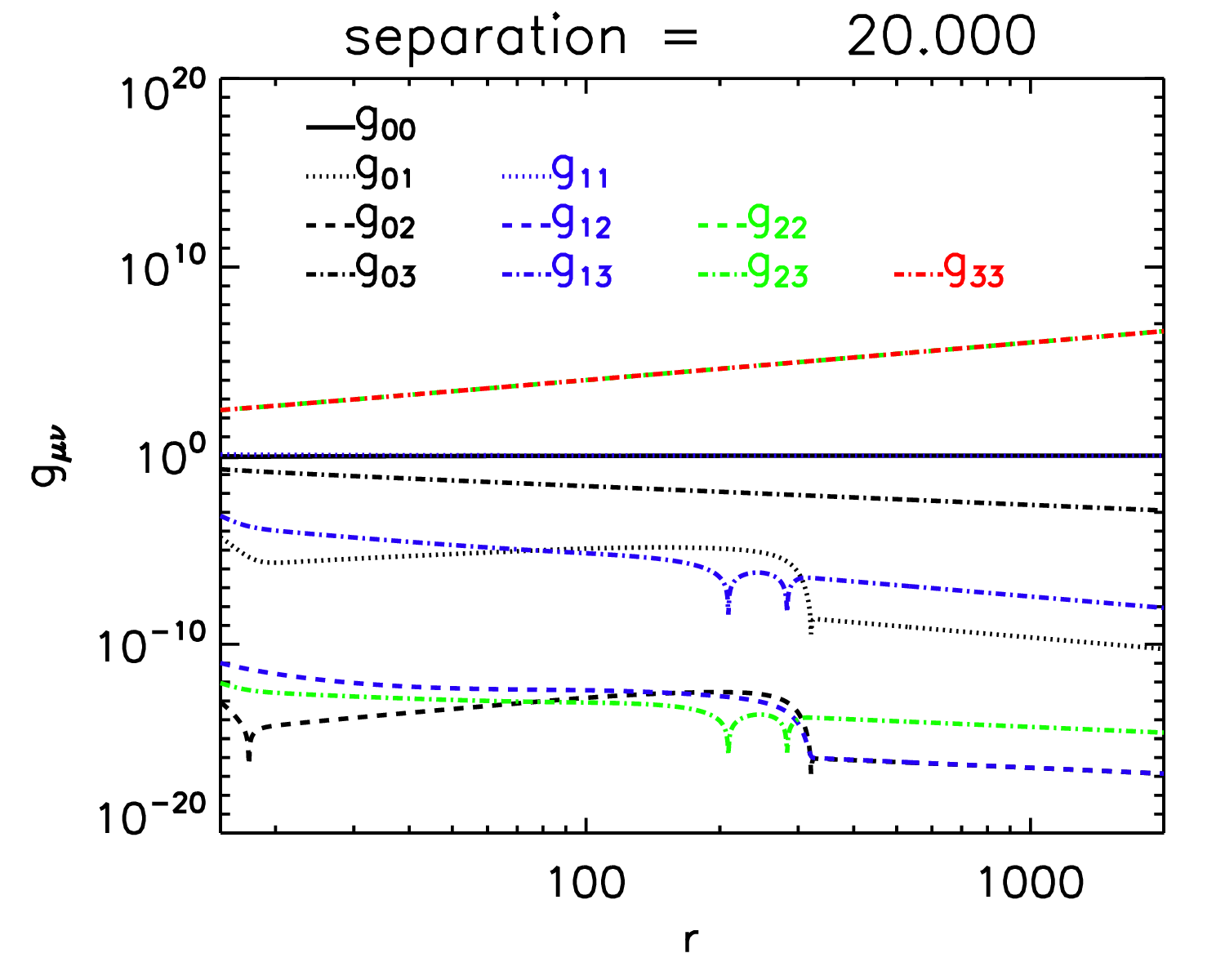}
  \includegraphics[width=0.4\textwidth]{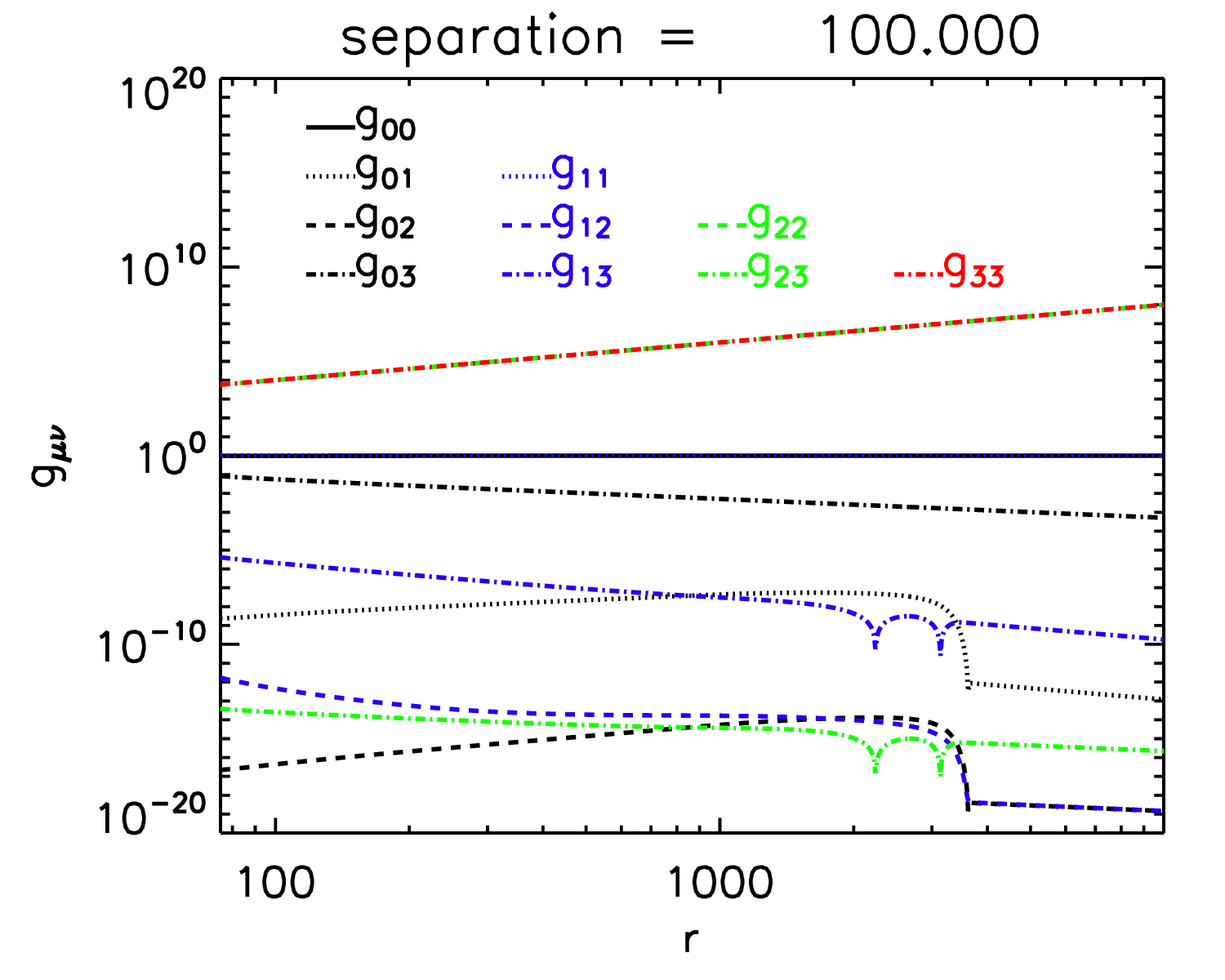}\\ 
  \includegraphics[width=0.4\textwidth]{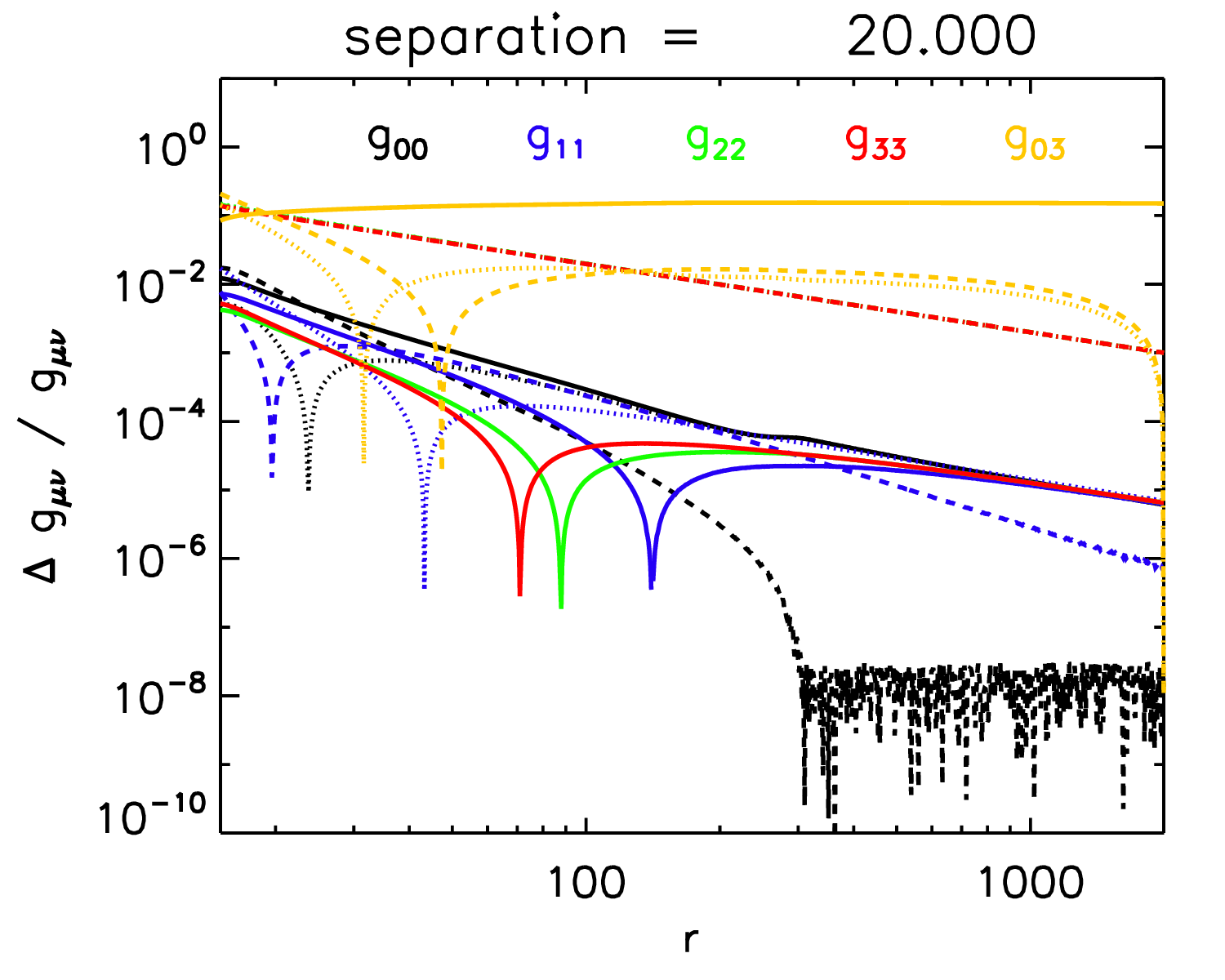}
  \includegraphics[width=0.4\textwidth]{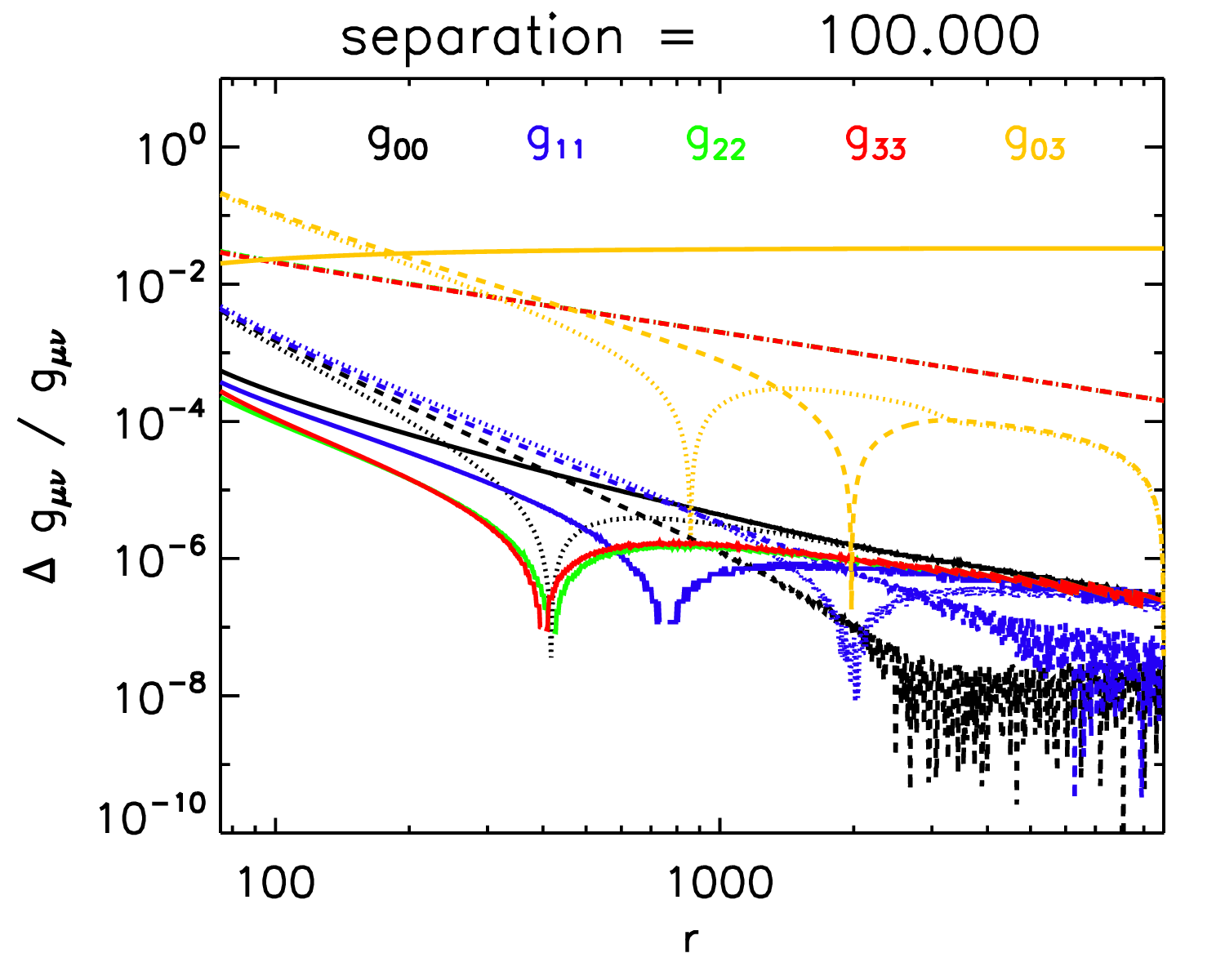} 
  \caption[]{Covariant metric components of the time-averaged metric used for 
    constructing disk initial data, for $a=20M$ (left column) and $a=100M$ (right column) binary separations, 
    and the 1PN metric (top row) and 2.5PN metric (middle row).   The relative differences (bottom row) 
    are shown between between the 2.5PN metric and 1PN metric (solid), the 2.5PN metric and its ``closest''
    Kerr solution metric (dots), and the 1PN metric and its ``closest'' Kerr solution metric (dashes).  
    Note that all off-diagonal components, except for $g_{t\phi}$, 
    are very small. This is the crucial ingredient for our disk initial data 
    construction procedure. Here, the numerical indices $\{0,1,2,3\}$ refer to the 
    coordinates $\{t,r,\theta,\phi\}$, respectively.
    \label{fig:a20_a100_t0} }
\end{figure*}

Our time-dependent metric does not admit any stationary disk
solutions. However, far from the binary, where the gas timescale is
much longer than the binary orbital period, we expect the disk to
behave as if it were evolving on  an effectively $\phi$-averaged spacetime.
If we hold the binary separation constant, this $\phi$-averaged
metric is stationary and therefore admits stationary disk
configurations as well. We therefore generate our {\it
quasi-equilibrium} initial data by finding stationary disks about
a $\phi$-averaged spacetime. We note that this $\phi$-averaged
spacetime is only used to generate initial disk configurations, not for
subsequent evolutions.
 For details on this
procedure, we refer the reader to Appendix~A of~\cite{Noble:2012xz}.
Our procedure for generating the initial disk configuration assumes
that all off-diagonal components (except for $g_{t\phi}$) vanish. In
practice, these are very small, but become larger as the 
binary separations is reduced.
We illustrate this in Fig.~\ref{fig:a20_a100_t0}, where we show for 
$a=20M$ and $a=100M$ binary separations plots of metric components for 
both 1PN and 2.5PN metrics, as well as the relative differences between 
them and the Kerr metric in Boyer-Lindquist coordinates with same total 
mass.  The differences in the $g_{t\phi}$ component of the 1PN and 2.5PN metrics 
can be approximated as differences in their dipole moments or ``spins.'' 
For instance, in Boyer-Lindquist coordinates, $g_{t\phi}\simeq 2a/r$ 
for $r\gg M$.  We find that the azimuthally-averaged 1PN and 2.5PN $g_{t\phi}$  components have the same functional form, 
but have different values for the spin parameter $a$ (at the level of a few percent).  
We fit an angle-averaged PN metric to this formula at our grid's outermost
radius to arrive at a spin parameter $a$ for the given PN metric; this parameter is then used to evaluate the Boyer-Lindquist form 
of the metric that we ultimately use for comparison purposes.  

For each separation, we make sure that disk solutions
have the same scale height, thus keeping the disks as similar as possible.


\subsection{Evolution equations}
\label{sec:evolution-eqs}

We solve the MHD equations with the \harm code~\cite{Noble:2008tm}.
As in~\cite{Noble:2012xz}, we assume that the gas does not self-gravitate and alter the spacetime
dynamics. We therefore need only solve the GRMHD equations on a specified background
spacetime, $g_{\mu \nu}(x^\lambda)$. 

The equations of motion originate from the local
conservation of baryon number density, the local conservation of stress-energy, and the
induction equations from Maxwell's equations (please see~\cite{Noble:2008tm} for more
details).  They take the form of a set of conservation laws:
\begin{equation}
\del_t \bU\left(\prim\right) = 
-\del_i \bF^i\left(\prim\right) + \mathbf{S}\left(\prim\right) \, 
\label{conservative-eq}
\end{equation}
where $\bU$ is a vector of ``conserved'' variables, $\bF^i$ are the fluxes, 
and $\mathbf{S}$ is a vector of source terms.  Explicitly, these 
are 
\begin{align}
\bU\left(\prim\right) & = \sqrt{-g} \left[ \rho u^t ,\, {T^t}_t 
+ \rho u^t ,\, {T^t}_j ,\, B^k  \right]^T \label{cons-U} \\
\bF^i\left(\prim\right) & = \sqrt{-g} \left[ \rho u^i ,\, {T^i}_t + \rho u^i ,\, {T^i}_j ,\, 
\left(b^i u^k - b^k u^i\right) \right]^T \label{cons-flux} \\
\mathbf{S}\left(\prim\right) & = \sqrt{-g} 
\left[ 0 ,\, 
{T^\kappa}_\lambda {\Gamma^\lambda}_{t \kappa}  - \mathcal{F}_t  ,\, 
{T^\kappa}_\lambda {\Gamma^\lambda}_{j \kappa}  - \mathcal{F}_j  ,\, 
0 \right]^T \, \label{cons-source}
\end{align}
where $g$ is the determinant of the metric, ${\Gamma^\lambda}_{\mu \kappa}$ are
the Christoffel symbols, $B^\mu = \dF^{\mu t}/\sqrt{4\pi}$ is our magnetic
field (proportional to the field measured by observers traveling orthogonal to the
spacelike hypersurface), $\dF^{\mu \nu}$ is the Maxwell tensor, $u^\mu$ is the
fluid's $4$-velocity, $b^\mu = \frac{1}{u^t} \left({\delta^\mu}_{\nu} + u^\mu
  u_\nu\right) B^\nu$ is the magnetic $4$-vector or the magnetic field projected
into the fluid's co-moving frame, and $W = u^t / \sqrt{-g^{tt}}$ is the fluid's
Lorentz function.  The MHD stress-energy tensor, $T_{\mu \nu}$, is defined as
\begin{equation}
T_{\mu \nu} = \left( \rho h + \bsq \right) u_\mu u_\nu   + \left( p + \bsq / 2\right) g_{\mu \nu} - b_\mu b_\nu \label{mhd-stress-tensor}
\end{equation}
where $\bsq \equiv b^\mu b_\mu$ is the magnetic energy density, $p$ is the gas pressure,
$\rho$ is the rest-mass density, $h = 1 + \epsilon + p/\rho$ is the specific
enthalpy, and $\epsilon$ is the specific internal energy.  The accretion flow is cooled to keep it close to a 
constant aspect ratio by removing excess heat to a radiation field, specified here as a 
radiative flux,  $\mathcal{F}_\mu = \mathcal{L}_c u_\mu$, with $\mathcal{L}_c$ being the fluid-frame cooling rate. 


We make use of piecewise parabolic reconstruction of the primitive variables at each cell interface for calculating the local Lax-Friedrichs flux~\cite{GMT03}, and a 
3-d version of the FluxCT algorithm is used to impose the solenoidal constraint,
$\partial_i \sqrt{-g} B^i = 0$~\cite{2000JCoPh.161..605T}.  The EMFs
(electromotive forces) are calculated midway along each cell edge using
piecewise parabolic interpolation of the fluxes from the induction equation~\cite{Zilhao:2013dta}.  A
second-order accurate Runge-Kutta method is used to integrate the equations of motion using the
method of lines once the numerical fluxes are found.  The primitive variables
are found from the conserved variables using the ``2D'' scheme of~\cite{Noble06}.  
Please see~\cite{Noble:2008tm} for more details.

\section{Hydrodynamic evolutions}
\label{sec:hydro}

Before embarking on full 3D MHD explorations, we performed 2D
equatorial evolutions of inviscid hydrodynamic (non-magnetized) disks
on  the background BBH spacetime for different PN orders.  We
initialized the disk using the procedure mentioned in
Sec.~\ref{sec:techniques} (and outlined in detail in Appendix~A of
\cite{Noble:2012xz}), where we set the aspect ratio to $H/r \simeq
0.1$.  In order to simplify the subsequent analysis, we have artificially
kept the binary separation fixed at $a/M=100, 50, 40, 30, 20, 15$.  We
denote the simulations by 
\hydroonepn{XX} and
\hydrotwopn{XX} for the 
1PN and 2.5PN cases respectively, where
XX is the binary separation in units of $M$.

For all the 2D simulations presented here, the
computational domain consisted of $320\times320$ cells with an outer
boundary at $R_{\rm out}=15a$  and an inner boundary at $R_{\rm
in}=0.75 a$.  On top of this computational domain, we constructed two
different types of disk configurations.  For the larger binary
separations of $a/M=100$, $a/M=50$, and $a/M=40$, we set up the disk
such that its inner edge was located at $r_{\rm in}=2.5a$ with the
radius of the pressure maximum at $r_{\rm pmax}=4.2a$, while for $a/M
\leq 40$ we set these to $r_{\rm in}=3a$ and $r_{\rm pmax}=5a$
respectively. Note that for $a/M=40$ we performed evolutions with
both configurations.

Here we analyze the effects of PN order on the disk
evolution by examining its influence on the disk's surface density,
torque density, and the mass enclosed within specified radii.

\subsection{Torque density}
\label{sec:torque}

\begin{figure*}[tbp]
  \centering
  \includegraphics[width=0.45\textwidth]{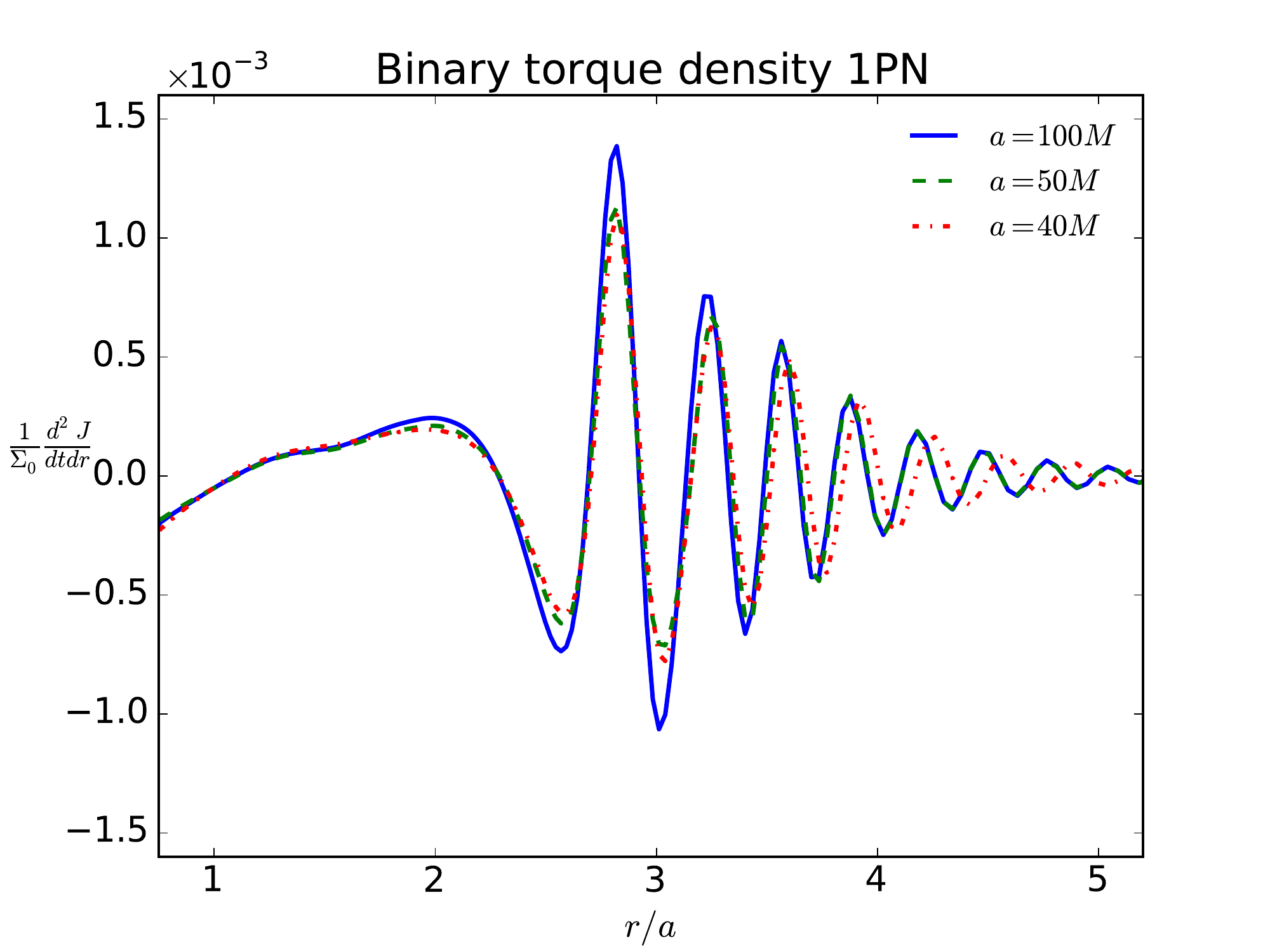}
  \includegraphics[width=0.45\textwidth]{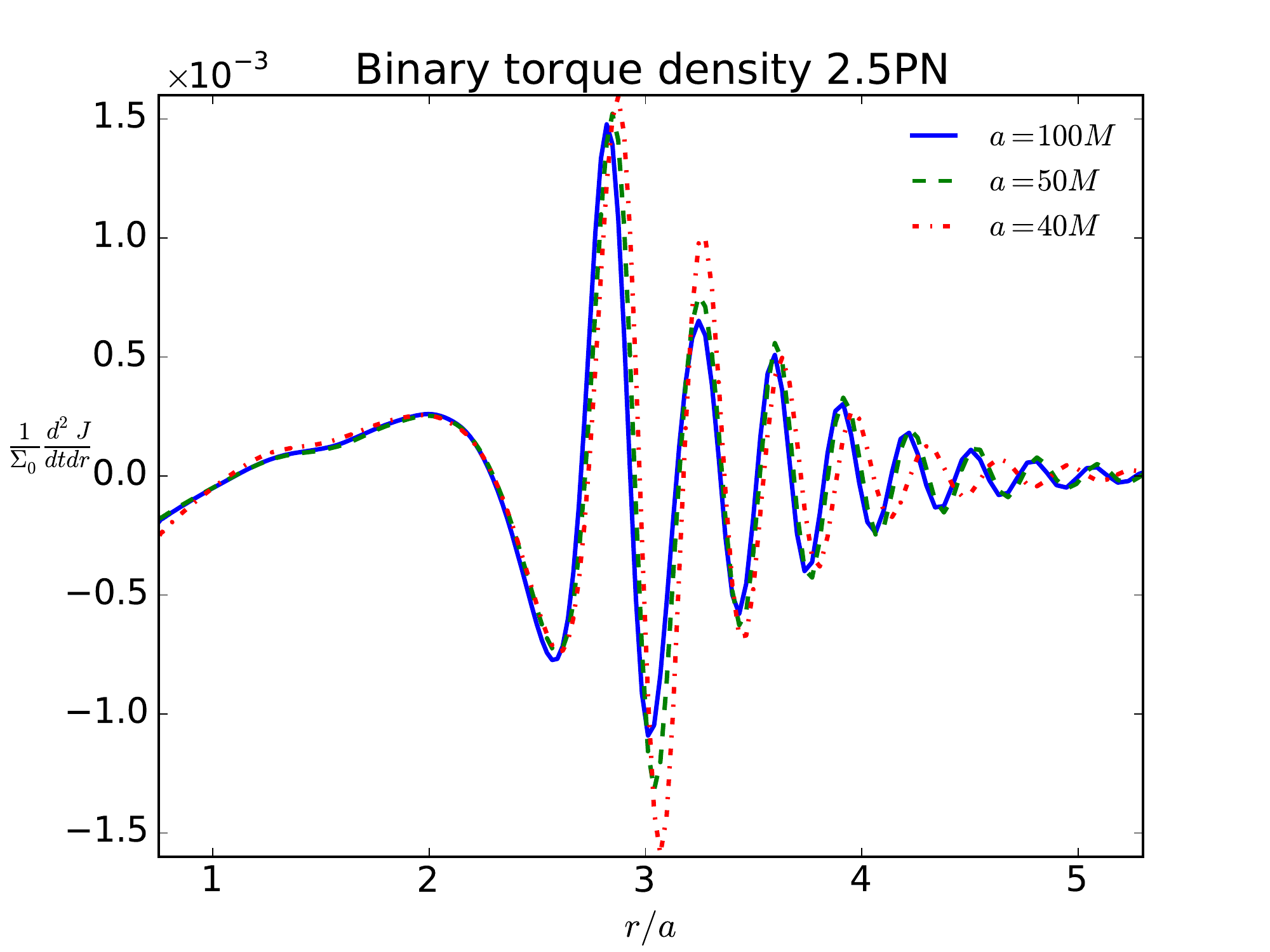}
  \caption[]{Torque density per unit radius due to binary potential for
    different separations. Quantities were time-averaged over the ``quasi-steady
    state'' period. \label{fig:stress-a100-40} }
\end{figure*}

\begin{figure*}[tbp]
  \centering
  \includegraphics[width=0.45\textwidth]{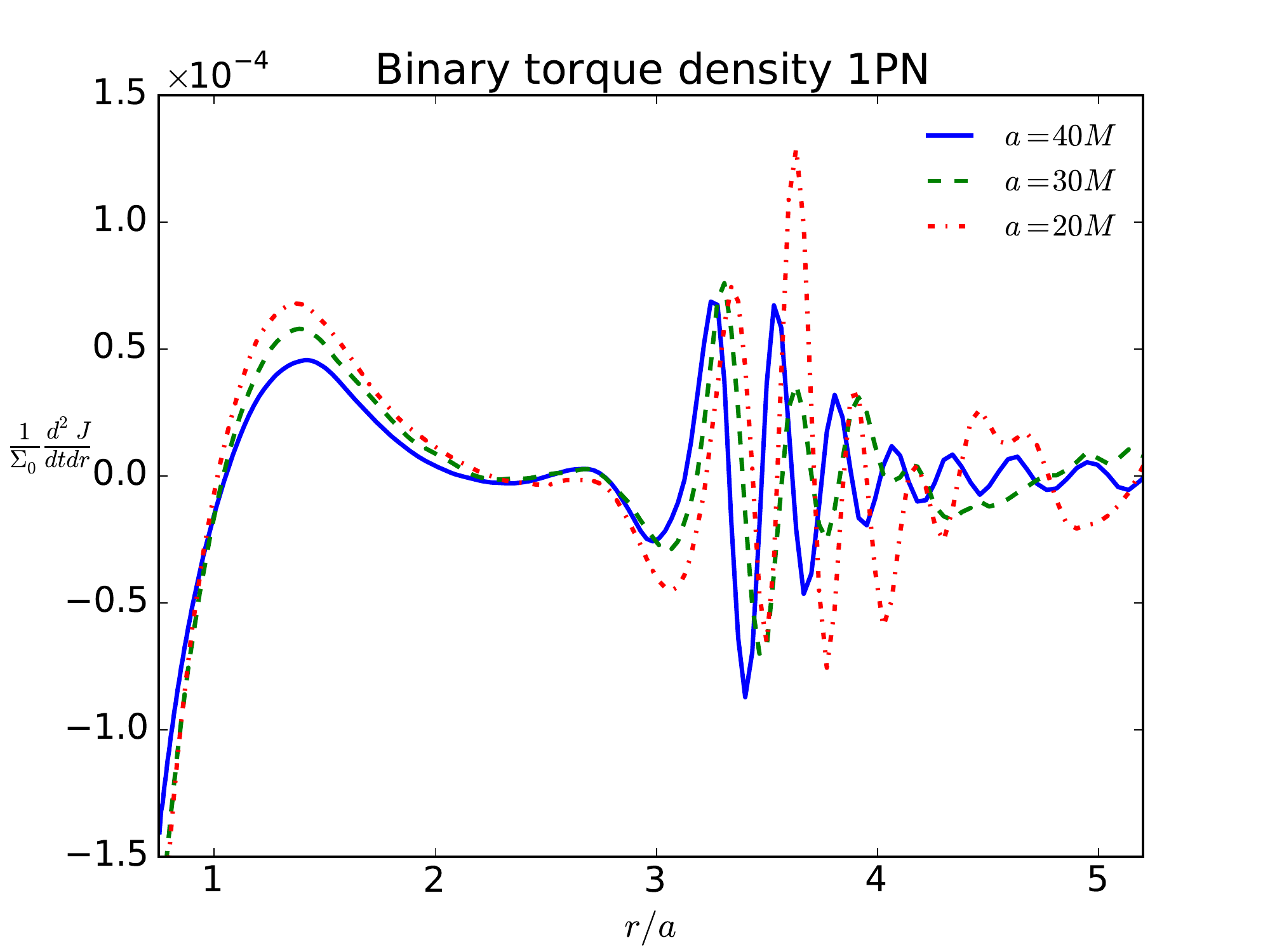}
  \includegraphics[width=0.45\textwidth]{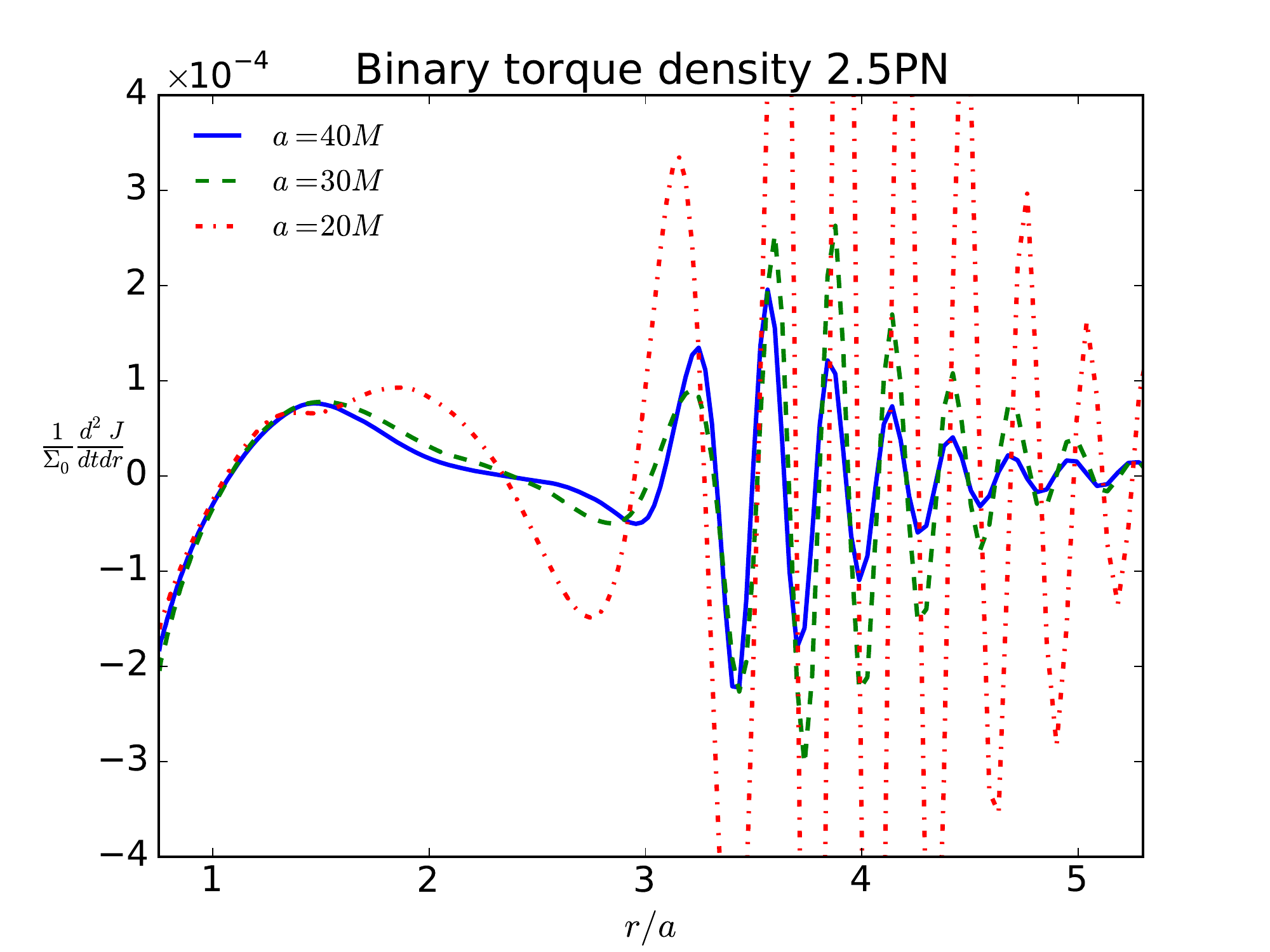}
  \caption[]{Same as Fig.~\ref{fig:stress-a100-40}, but for separations $a/M=40,
    30, 20$. Recall that the disk was initialized with a different configuration
    than that of Fig.~\ref{fig:stress-a100-40}. Note that the $y$-axis range of
    these figures is different, as the finer details of the right-hand side
    figure would not be noticeable otherwise. \label{fig:stress-a40-15} }
\end{figure*}

We found that one of the most sensitive measures of differences
 in the background
spacetime is the torque density.
The disk is torqued by the binary through the non-axisymmetric nature of the gravitational potential. In the Newtonian limit, the time-averaged torque density takes the form~\cite{Macfadyen:2006jx}
\begin{equation}
\label{eq:torque-density-newt}
\frac{dT}{dr} = -2\pi r \left \langle 
  \Sigma \frac{\partial \Phi}{\partial \phi}
\right \rangle \,,
\end{equation}
where $\Phi$ is the Newtonian gravitational potential of the binary, and where 
$\langle X \rangle \equiv  \frac{\int X \, \sqrt{-g} \, d\theta d\phi}{\int \sqrt{-g} \, d\theta d\phi}$ 
denotes the average over spheres. 
From this formula, it is immediately apparent that the more the gravitational potential deviates from axisymmetry, the greater the corresponding communicated torque density will be for a given asymmetric (and correlated) distribution of $\Sigma$.
We therefore expect that, for the same disk configuration, the torque density will be larger for our 2.5PN runs as the 
higher order PN terms give rise to larger asymmetries when the terms become important. 

For completeness, we also write the corresponding general relativistic torque density formula (which is the one we actually compute)~\cite{Farris:2011vx}
\begin{equation}
\label{eq:torque-density-GR}
\frac{dT}{dr} = \int \sqrt{-g} T^{\mu}{}_{\nu} \left( \nabla_{\mu} \phi^{\nu} \right)  \, r \,dz \, d\phi \,,
\end{equation}
where $\phi^{\mu} \equiv (\partial_{\phi})^{\mu}$.

In Figs.~\ref{fig:stress-a100-40} and \ref{fig:stress-a40-15}, we plot
the time-averaged binary torque density for our 1PN and 2.5PN
simulations for $a/M=100, 50, 40$ and $a/M=40, 30 ,20$. Here the
average angular momentum flux is calculated by integrating from orbit
120 to orbit 240, which is safely in the ``quasi-steady state''
regime.

In all cases, the binary torque density $dT/dr$ is strongest for
$a\lesssim r \lesssim 5a$ and shows a strong correlation with the
Reynolds stresses.  Note how in Fig.~\ref{fig:stress-a100-40} all
curves exhibit the same basic decaying oscillatory behavior with very
similar amplitudes, as is expected in the quasi-Newtonian regime (in
the Newtonian limit, all plots would overlap).  The oscillatory
pattern observed is a standard feature of such hydrodynamic
evolutions, and is due to the presence of spiral density waves in the
inner disk cavity~\cite{Macfadyen:2006jx}. We see that for $a/M > 40$,
the 1PN curves lie on top of each other at larger radii, which is
expected in the Newtonian regime. The $a/M=40$ curve has a slightly
larger wavelength. The corresponding 2.5PN curves show a similar
behavior but there is also a noticeable difference in wavelength
between the $a/M=100$ and $a/M=50$ curves. Comparing the 1PN and 2.5PN
curves for $a/M\geq40$, we see that at smaller $a/M$ the 2.5PN curves
show a larger amplitude than the 1PN curves. This latter effect is
relatively small but becomes much stronger at smaller $a/M$. In
Fig.~\ref{fig:stress-a40-15}, we show the torque density for $a/M =
40, 20,15$ with a slightly different disk configuration. With this new
disk configuration, the differences in amplitude between 1PN and 2.5PN at $a/M=40$
are magnified. Most importantly, the trends in amplitude with 
decreasing $a/M$ are completely different between 1PN and 2.5PN. While 1PN shows a decrease
in amplitude from $a/M=40$ to $a/M=30$, 2.5PN shows a strong increase.
Note the relative scales in the figure. At $a/M=20$ the 1PN and 2.5PN
curves are no longer even qualitatively similar.
It thus appears that one needs to include 2.5PN corrections to the
metric even at separations as large as $a/M=30$.



\subsection{Surface density}
\label{sec:surf-dens}

We define the surface density $\Sigma$ as
\begin{equation}
  \Sigma(t,r,\phi) \equiv \int  \rho  \sqrt{-g} dz
  \label{eq:sigma}
\end{equation}
and $\Sigma(t,r)$ will denote the azimuthal average of the above equation. 
For later convenience, we define $\Sigma_0$ to be the maximum value
for the surface density at $t=0$.
This quantity will be useful for normalization purposes.

\begin{figure*}[tbp]
  \centering
  \includegraphics[width=0.45\textwidth]{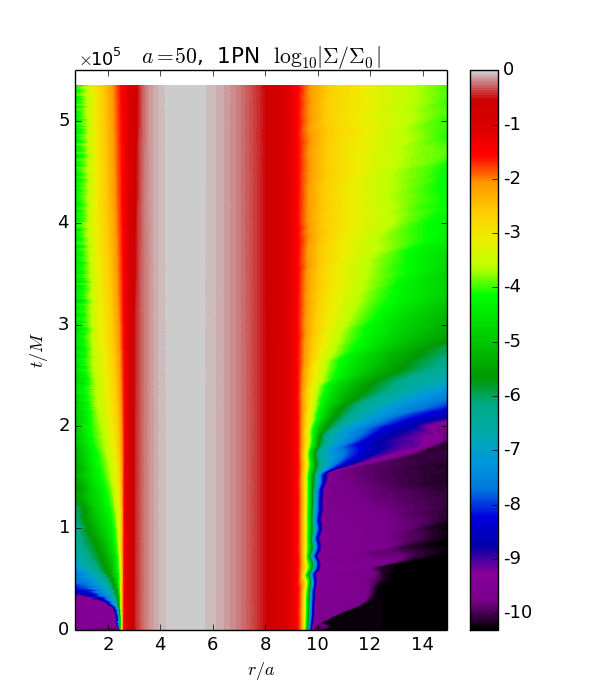} 
  \includegraphics[width=0.45\textwidth]{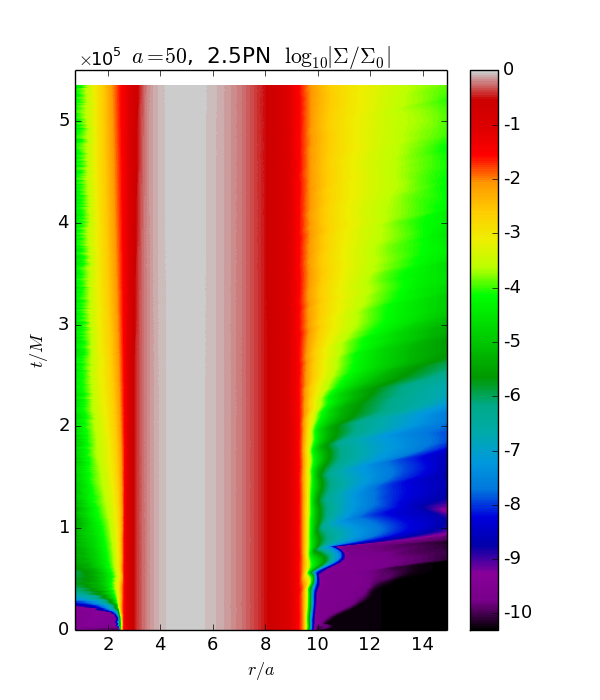} \\
  \includegraphics[width=0.45\textwidth]{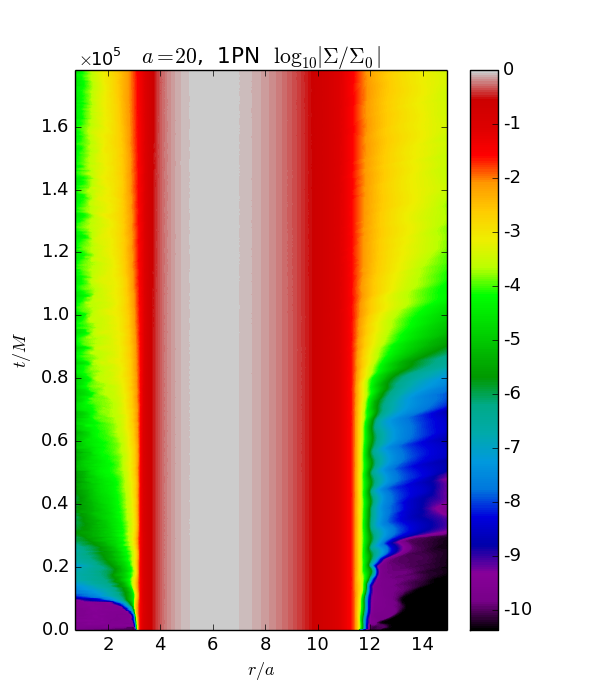} 
  \includegraphics[width=0.45\textwidth]{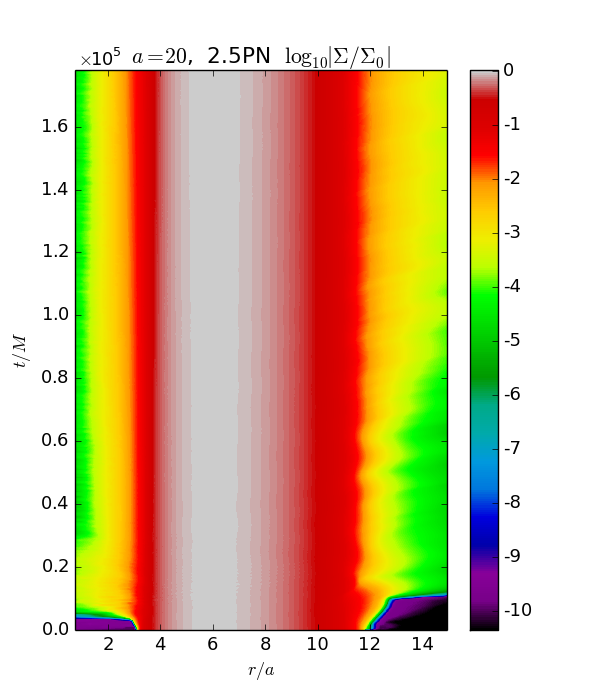}
  \caption[]{Color contour of surface density
    $\Sigma(t,r)$, Eq.~\eqref{eq:sigma-rel-diff} (in logarithmic units) for
    some simulations. 
    $y$-axis is the simulation time in units of total mass $M$ and $x$-axis
    is the coordinate radial distance in units of binary separation
    $a$. \label{fig:pn_log_sigma_r_t} }
\end{figure*}

In Fig.~\ref{fig:pn_log_sigma_r_t} 
we show
the behavior of the surface density (in logarithmic units) for various binary
separations and PN orders. These
plots 
give us a picture of how the gap fills in and how the gas diffuses out of the
disk. At $a/M=50$, the rate and amount of gas filling in the gap and the rate
and amount of gas diffusing out of the disk are essentially identical
between 1PN and 2.5PN.
Even at
$a/M=20$, the final distribution of surface density is again independent of PN order. Here,
however, there is a considerable lag in the time it takes for gas to start
diffusing out of the disk at 1PN. Interestingly, while the torque
density for 1PN and 2.5PN are quite different at $a/M=20$,
the net effect seems to be only to accelerate the equilibration of the
2.5PN disk---the resultant quasi-equilibrium state of the 1PN and 2.5PN  
disks are largely unaffected by the differences in these torques.

\begin{figure*}[tbhp]
  \centering
  \includegraphics[width=0.45\textwidth]{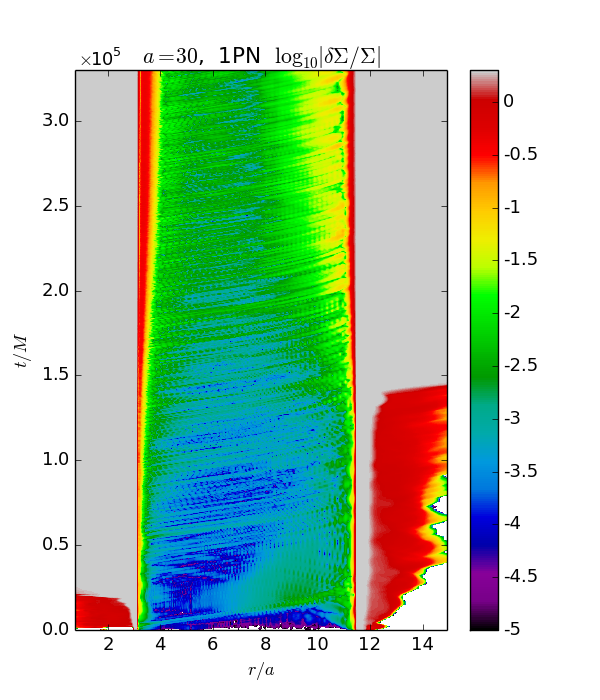} 
  \includegraphics[width=0.45\textwidth]{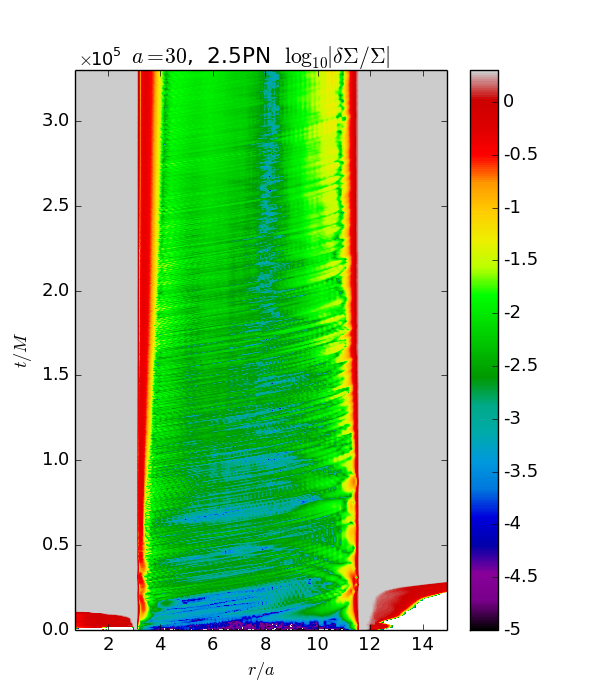} \\
  \includegraphics[width=0.45\textwidth]{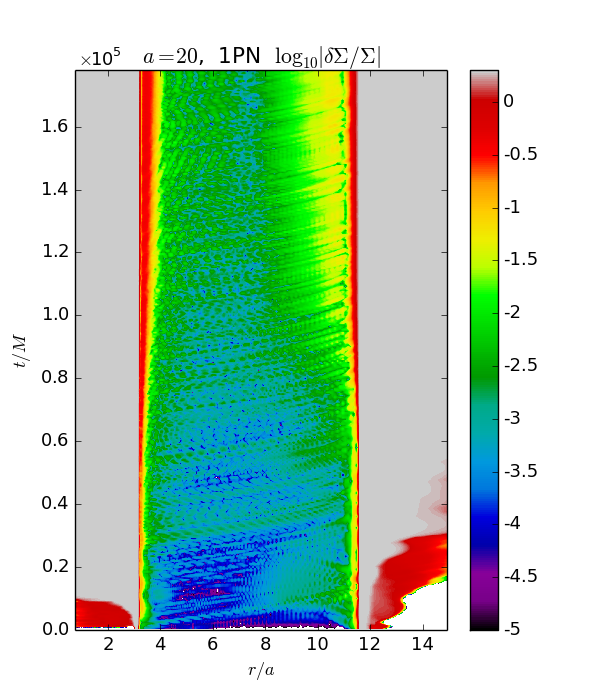} 
  \includegraphics[width=0.45\textwidth]{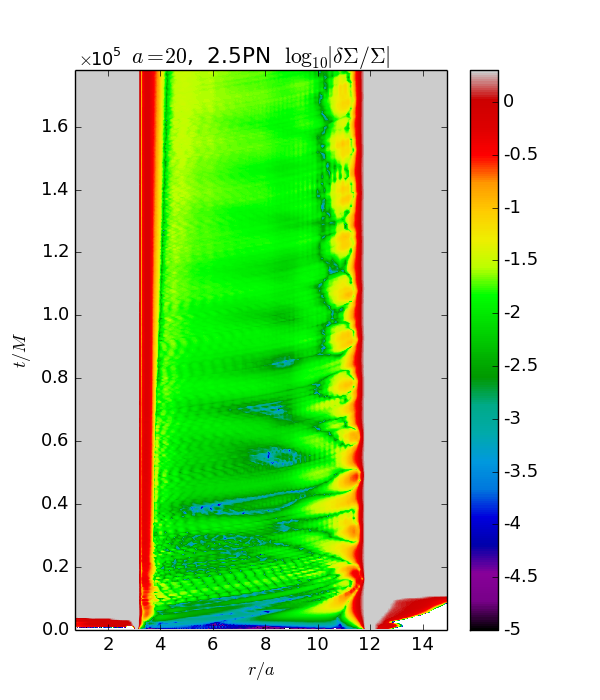}
  \caption[]{Color contour of (relative difference of) surface density
    $\Sigma(t,r)$, Eq.~\eqref{eq:sigma-rel-diff} (in logarithmic units) for
    simulations \hydroonepn{30}, \hydrotwopn{30}, \hydroonepn{20}, and
    \hydrotwopn{20}. $y$-axis is the simulation time in units of total mass $M$
    and $x$-axis is the coordinate radial distance in units of binary separation
    $a$. \label{fig:delta_sigma_vs_r_t-a30-a20} }
\end{figure*}

To further evaluate the effects of PN order on the bulk of the disk,
we examine the relative difference between $\Sigma(t,r)$ and its
initial value $\Sigma(0,r)$: 
\begin{equation} \frac{\delta \Sigma}{\Sigma} \equiv 2 \left|
\frac{\Sigma(t,r) - \Sigma(0,r)}{|\Sigma(t,r)| + |\Sigma(0,r)|}
\right|.  \label{eq:sigma-rel-diff} \end{equation}
In
Fig.~\ref{fig:delta_sigma_vs_r_t-a30-a20} 
we plot this quantity for binary
separations of $a=20M$ and $a=30M$, which (for this particular
measure) are illustrative examples of all simulations performed.  
The relative change in $\Sigma$ is complementary to a plot of the $\Sigma$ itself since
it better illustrates the behavior of the bulk, rather than the
behavior of the inner and outer edges.  The relative change is off scale before and 
beyond the disk because the atmosphere was initialized with very little density.
This figure shows how far from equilibrium our
initial data is. Concentrating
on the bulk of the disk itself, we see very little difference between
the 1PN and 2.5PN disks at $a/M=30$. Both the 1PN and 2.5PN disks
settle down to configurations that are very close to their initial
configurations, with the largest deviations near the inner and outer
edges. At  $a/M=20$ some
differences are apparent. The 2.5PN disk is further from its initial
configuration at all radii (but the effect is small) than the 1PN disk, and the 2.5PN
disk exhibits decaying oscillatory modes near
the outer edge not present in the 1PN disk.
 While we  note that these oscillations may be due, in part,
to the construction of the initial data (which assumes zero
off-diagonal terms in the $\phi$-averaged metric),
it  seems to us that  the 2.5PN effects during evolution are
important for this measure of the behavior of the disk at
$a/M\sim20\text{--}30$.

We note also that in order to rule out the possibility that the observed differences between the 1PN and 2.5PN simulations be due to the accumulation of numerical errors over time, we have performed higher resolution versions of simulations \hydroonepn{20} and \hydrotwopn{20}.
See Appendix~\ref{app:details} for this discussion.

\subsection{Mass enclosed}
\label{sec:mass-enclosed}

\begin{figure*}[tbp]
  \centering
  \includegraphics[width=0.32\textwidth]{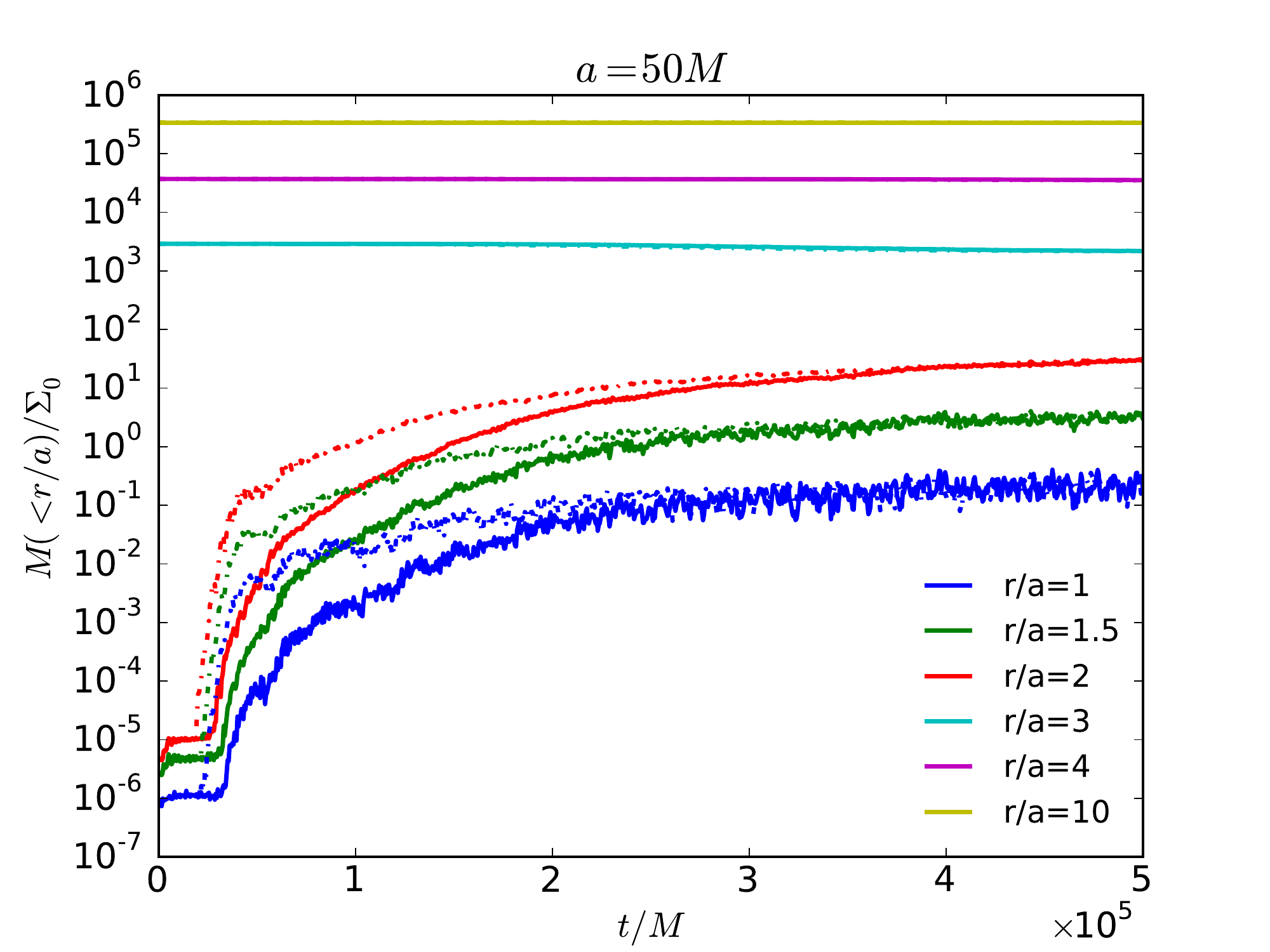} 
  \includegraphics[width=0.32\textwidth]{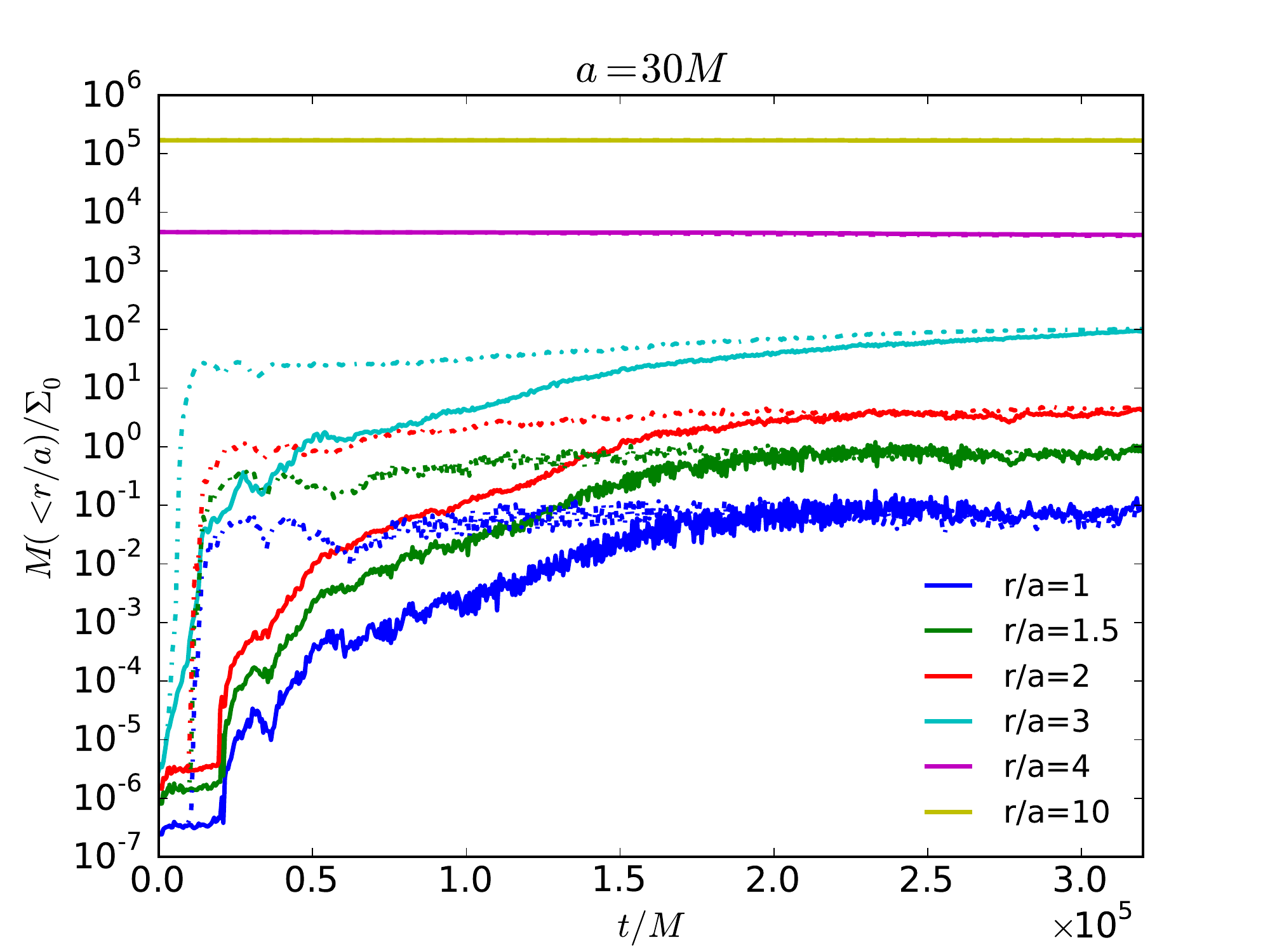}
  \includegraphics[width=0.32\textwidth]{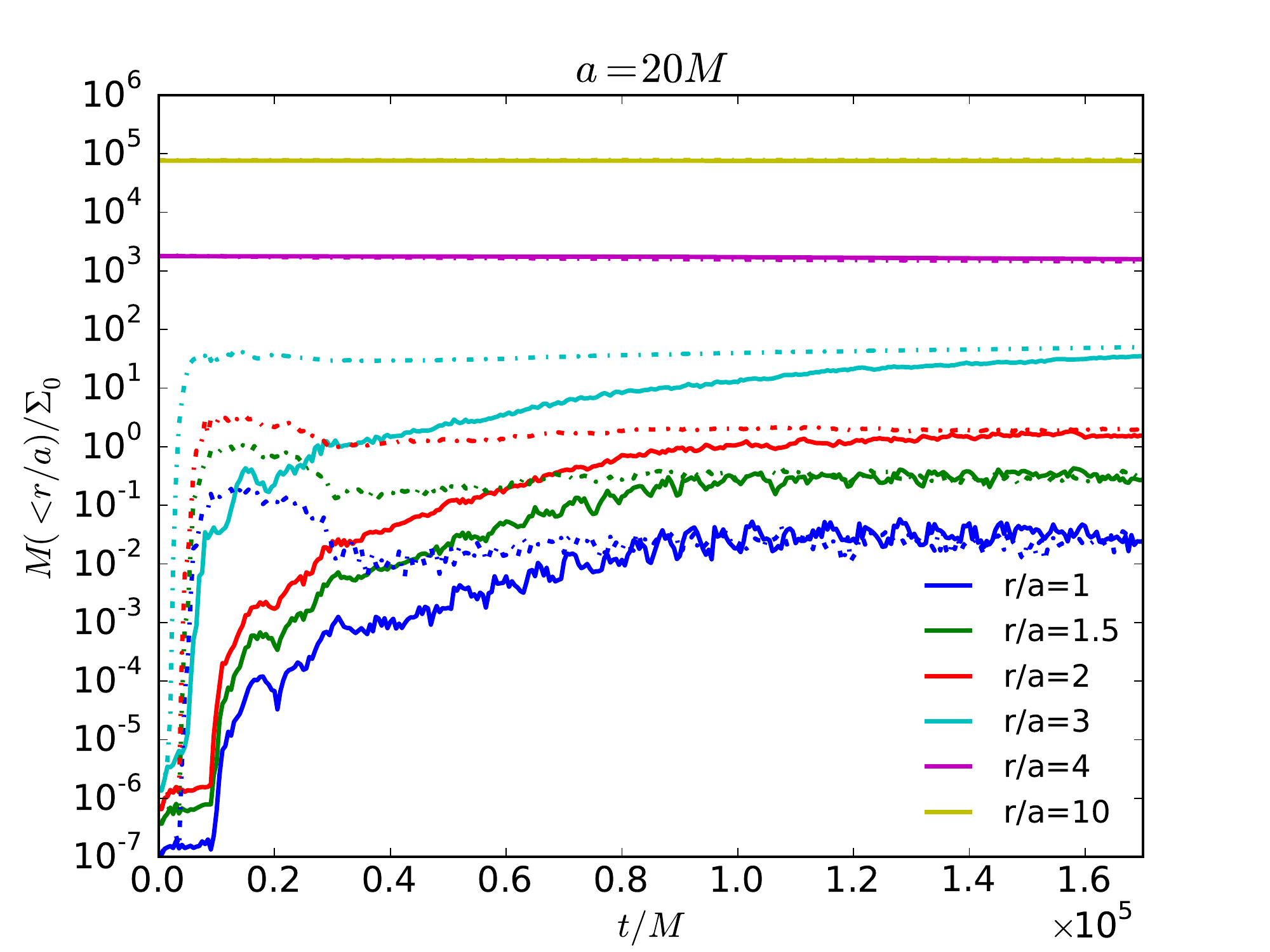} 
  \caption[]{Mass (of the gas) enclosed within several radii as functions of
    time. Full lines show 1PN evolutions while dotted lines show the
    corresponding 2.5PN versions.\label{fig:mass-enc-a50-20} }
\end{figure*}

We close this Section with a discussion of how 
the disk's mass flow equilibrium
changes over time.  Please see Fig.~\ref{fig:mass-enc-a50-20} 
for plots of the time series of the enclosed mass within spherical 
volumes of different radii. 
We begin by noting that for $a \gtrsim 50M$ no
significant difference can be seen between 1PN and 2.5PN
evolutions.  Evolutions for smaller binary separations show a
different picture, though.  Indeed, for all 1PN evolutions, the amount
of mass inside small radii increases monotonically with time, until it
saturates at late times.  For the corresponding 2.5PN evolutions with
binary separation $a \lesssim 40M$, however, the mass enclosed
exhibits a much more rapid growth, followed by a small decay.

One critical thing to note is that while the initial transient
behavior of the disks at smaller separations are quite different
between 1PN and 2.5PN (with 2.5PN equilibrating noticeably sooner), the
quasi-equilibrium state for 1PN and 2.5PN disks are very similar down to
separations as small as $a/M=20$. Apparently, 2.5PN seems to destabilize
the gas around the gap, allowing it to fall into the binary until
pressure inside the gap supports the inner edge of the disk. The
equilibrium state of the disk depends on the gas inside the gap,
therefore a configuration that leads to gas entering the gap sooner
can equilibrate faster.

Thus, it seems that even though the effects from PN spacetime order \emph{are} significant for
the smaller binary separations, hydrodynamic effects begin to dominate during the late
time stages of these evolutions.  Simulations with lower PN order merely need
more time to relax to the same configuration as their corresponding higher PN
order cases.  The apparent conclusion is that PN approximations do affect the
transient of the disk dynamics even at large separation ($a/M=30$) but the bulk
properties of the disk at later times are robust even at $a/M=20$.

\section{MHD evolutions}
\label{sec:mhd}

We now turn our attention to MHD simulations of circumbinary accretion
disks. Since 3D MHD simulations are much more computationally costly than (2D) hydrodynamic ones, we cannot afford to explore the parameter space as extensively as in the hydrodynamic case.
For our MHD runs, we have therefore chosen to fix the binary
separation at $a = 20M$ and only vary the PN order.

We prepared three different evolutions: the ``benchmark'' 2.5PN run (hereafter referred to as \mhdtwopn), a 1PN run where the disk was initialized with the same specific angular momentum at $r_{\rm in}$ as in the \mhdtwopn case (hereafter referred to as \mhdonepnold), and a 1PN run where the disk was chosen to have the same aspect ratio $r_{\rm pmax}$ as that of the \mhdtwopn case (hereafter referred to as \mhdonepnnew).  Because of the differences in the spacetime metric, it is impossible to find a disk with the same aspect ratio and specific angular momentum in the 1PN spacetime as was originally used in the 2.5PN 
spacetime. 

For our MHD simulations we used a 3D numerical grid with $300 \times 160
\times 400$ cells, an outer boundary at $R_{\rm out}=15a=300M$, and an inner
boundary at $R_{\rm in}=0.75 a = 15M$. The disk was chosen to have its inner
edge located at $r_{\rm in}=3a=60M$ with the radius of the pressure maximum at
$r_{\rm pmax}=5a=100M$. 
Please see Ref.~\cite{Noble:2012xz} for further details about the disk setup and parameters used, 
and Appendix~\ref{app:details} for a discussion about our use of the excision procedure and 
how well our runs resolve the magnetorotational instability (MRI). 

A detailed analysis of quantities and results pertaining to the
\mhdtwopn evolution was already performed in~\cite{Noble:2012xz}, to
where we refer the interested reader. We will therefore not repeat
this extensive analysis, and will focus our discussion here on
the differences between the 1PN and 2.5PN evolutions.   We find that each simulation enters a so-called ``secularly evolving'' state 
at around $t=40,000M$ in which many characteristics of the disk are seen to only gradually 
evolve in time; a circumbinary disk is not expected to be steady as the gravitational torque from the binary 
persists to perform work on the gas.  

\subsection{Torque}
\label{sec:mhd-torque}

\begin{figure*}[htbp]
  \centering
  \includegraphics[width=0.45\textwidth]{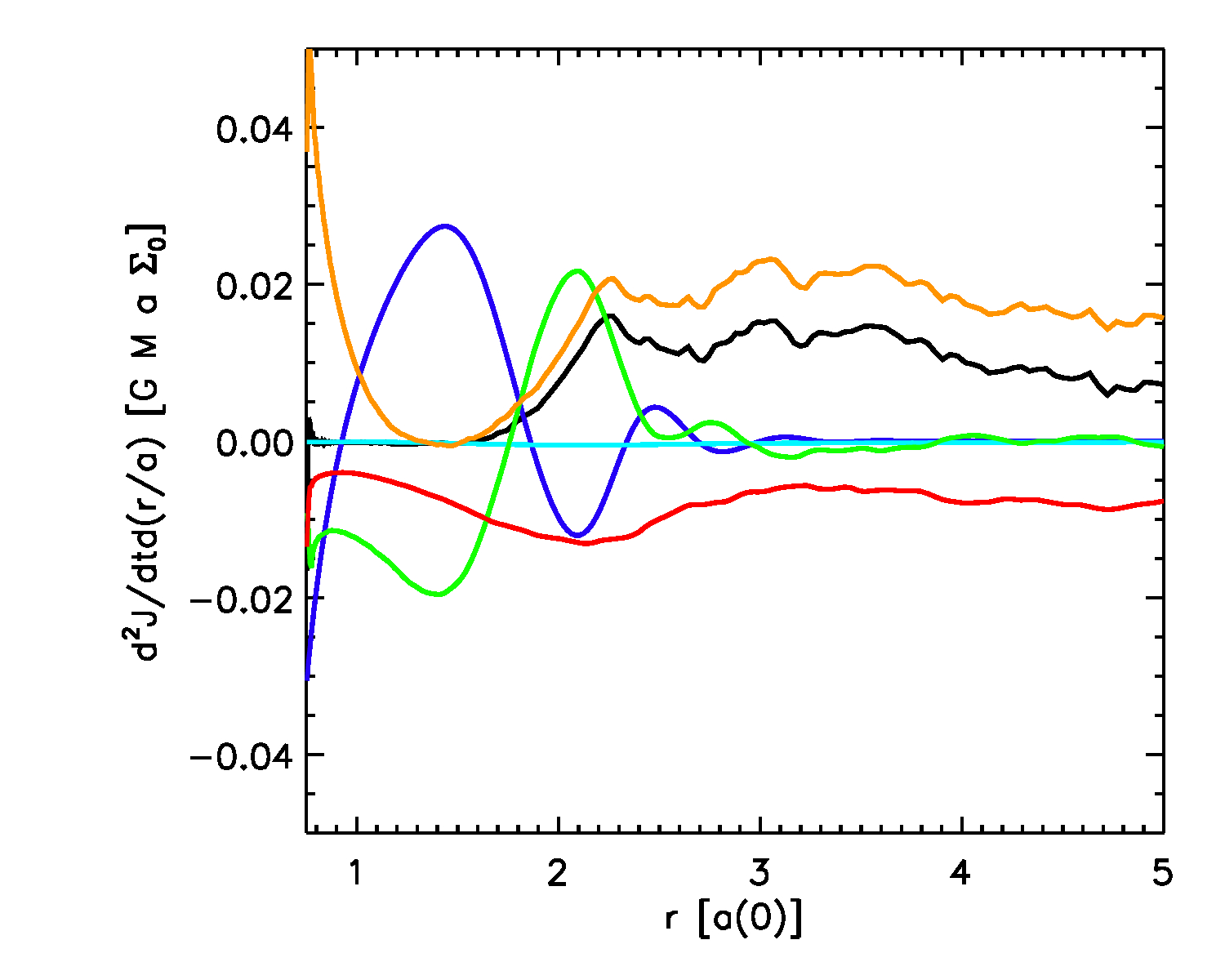}
  \includegraphics[width=0.45\textwidth]{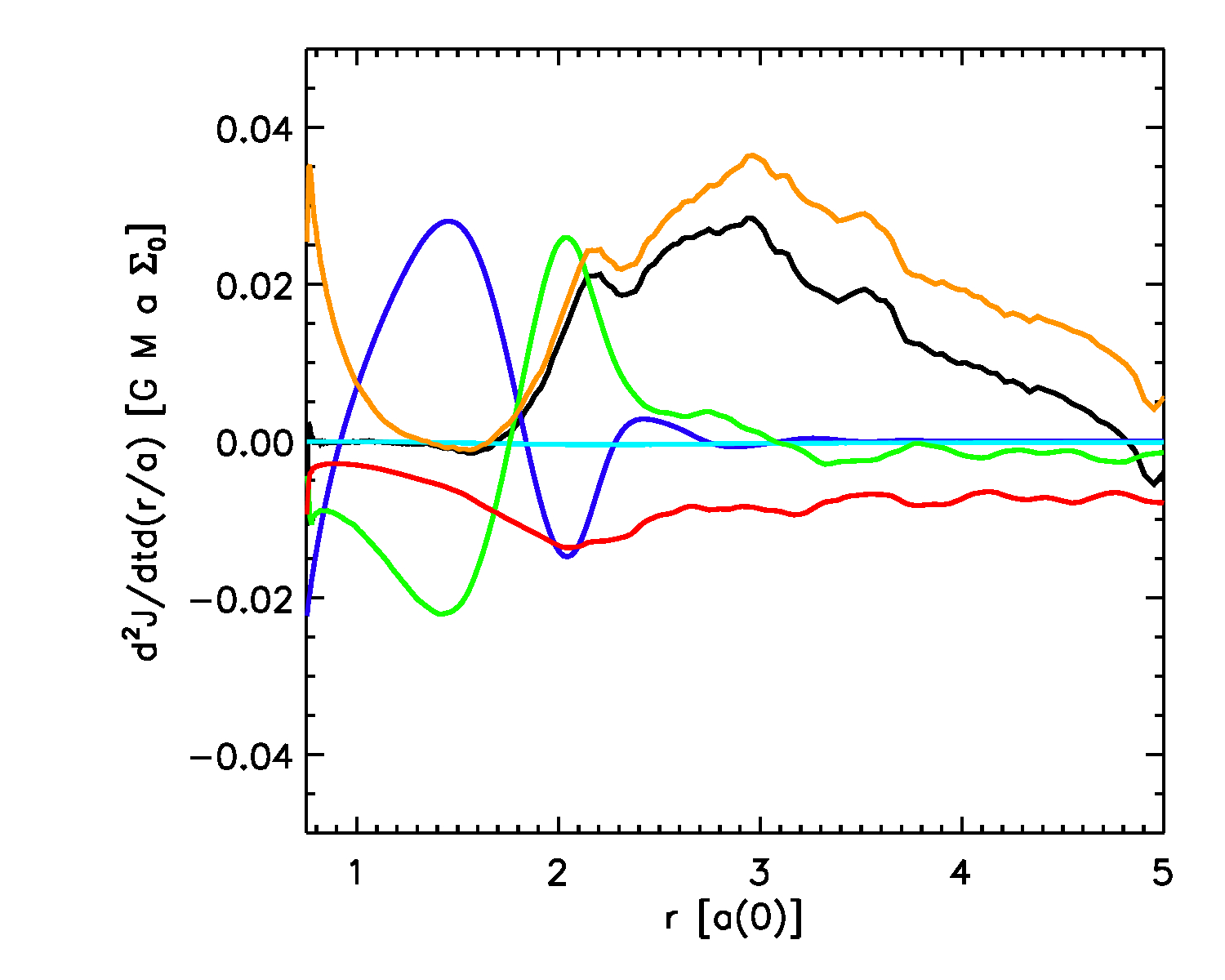}
  \caption[]{Different contributions to the flux of angular momentum through the accretion flow seen in the 
\mhdonepnnew (left) and \mhdtwopn (right) runs. 
Radial derivatives of the angular momentum flux due to
    shell-integrated Maxwell stress in the coordinate frame (red), the angular
    momentum flux due to shell-integrated Reynolds stress in the coordinate
    frame (green), advected angular momentum (gold), and net rate of change of
    angular momentum $\partial_r \partial_t J$ (solid black).  Also shown are 
    torque densities per unit radius due to the actual binary potential (blue) and radiation losses (cyan).
    All quantities are time-averaged over the secularly evolving period. 
    To clarify colors used, note that at $r=4a$
    colors are (from bottom to top): red, green, cyan, blue, black, gold.\label{fig:mhd-torque} }
\end{figure*}

We can see the angular momentum flow of the system in Fig.~\ref{fig:mhd-torque}, where we plot the radial derivatives of time-averaged angular momentum fluxes integrated on shells.
This is the plot analogous to Fig.~\ref{fig:stress-a40-15} for these MHD evolutions.
As expected, several differences stand out when comparing with the purely hydrodynamic evolutions, as in the hydrodynamic case there is no mechanism to efficiently transport angular momentum.
In this case, we see that the binary torque density $dT/dr$ is mostly delivered in the $a\lesssim r \lesssim 2a$ region. 
Most of the angular momentum is delivered in the gap, where the density of the fluid is much lower than in the disk proper.
As in the hydrodynamic case, a strong correlation with the Reynolds stresses is observed. 
And throughout the flow, Maxwell stress (not present in the hydrodynamic case) acts to remove angular momentum from the gas and carry it outward.

More important for our purposes here, though, is noting that runs \mhdonepnnew and \mhdtwopn show hardly any noticeable difference between them.
Thus, for this quantity (and unlike the corresponding hydrodynamic case), MHD dynamics seem to dominate over and mask ``spacetime'' related effects.

\subsection{Surface density}
\label{sec:mhd-surf-dens}

\begin{figure*}[tbhp]
  \centering
  \includegraphics[width=0.45\textwidth]{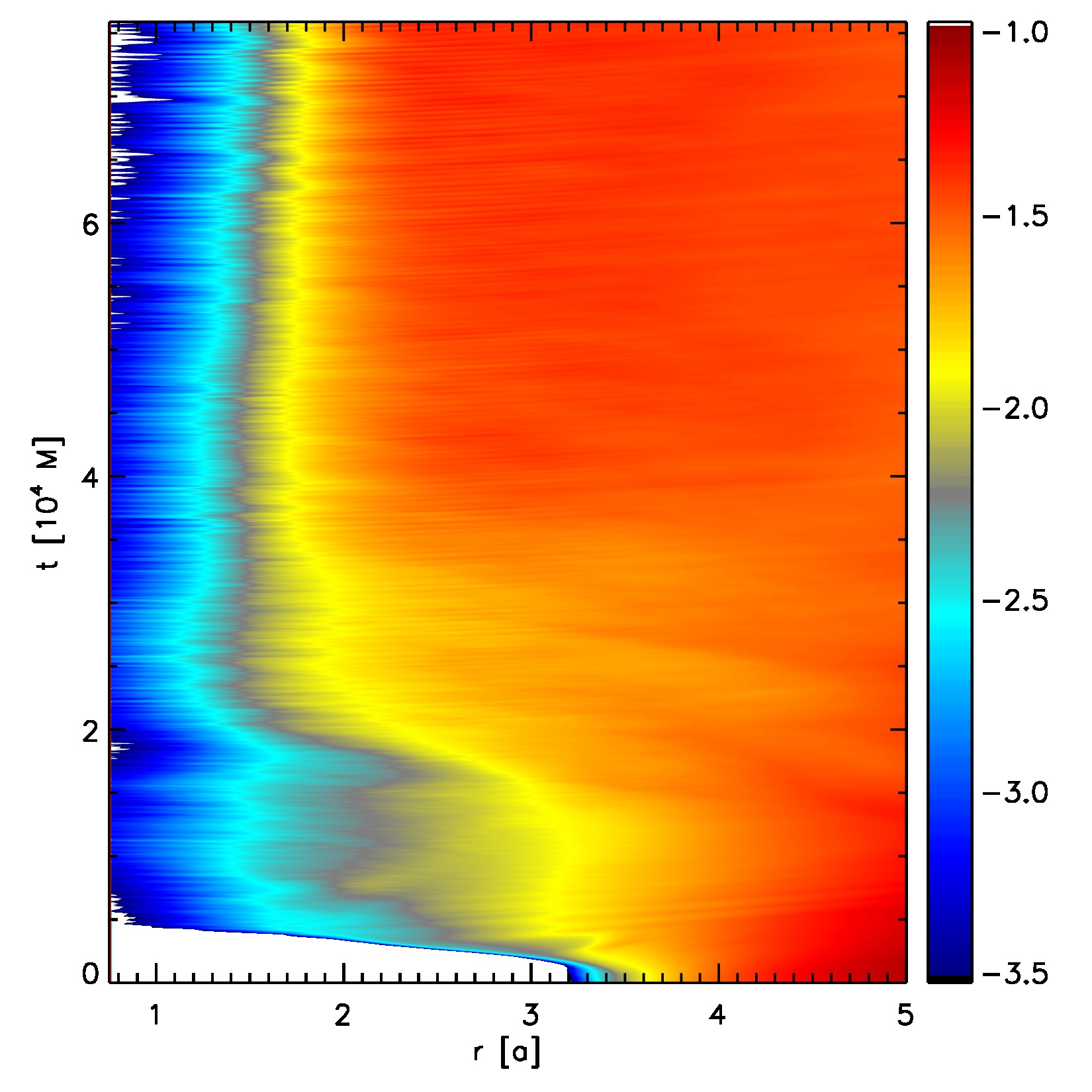}
  \includegraphics[width=0.45\textwidth]{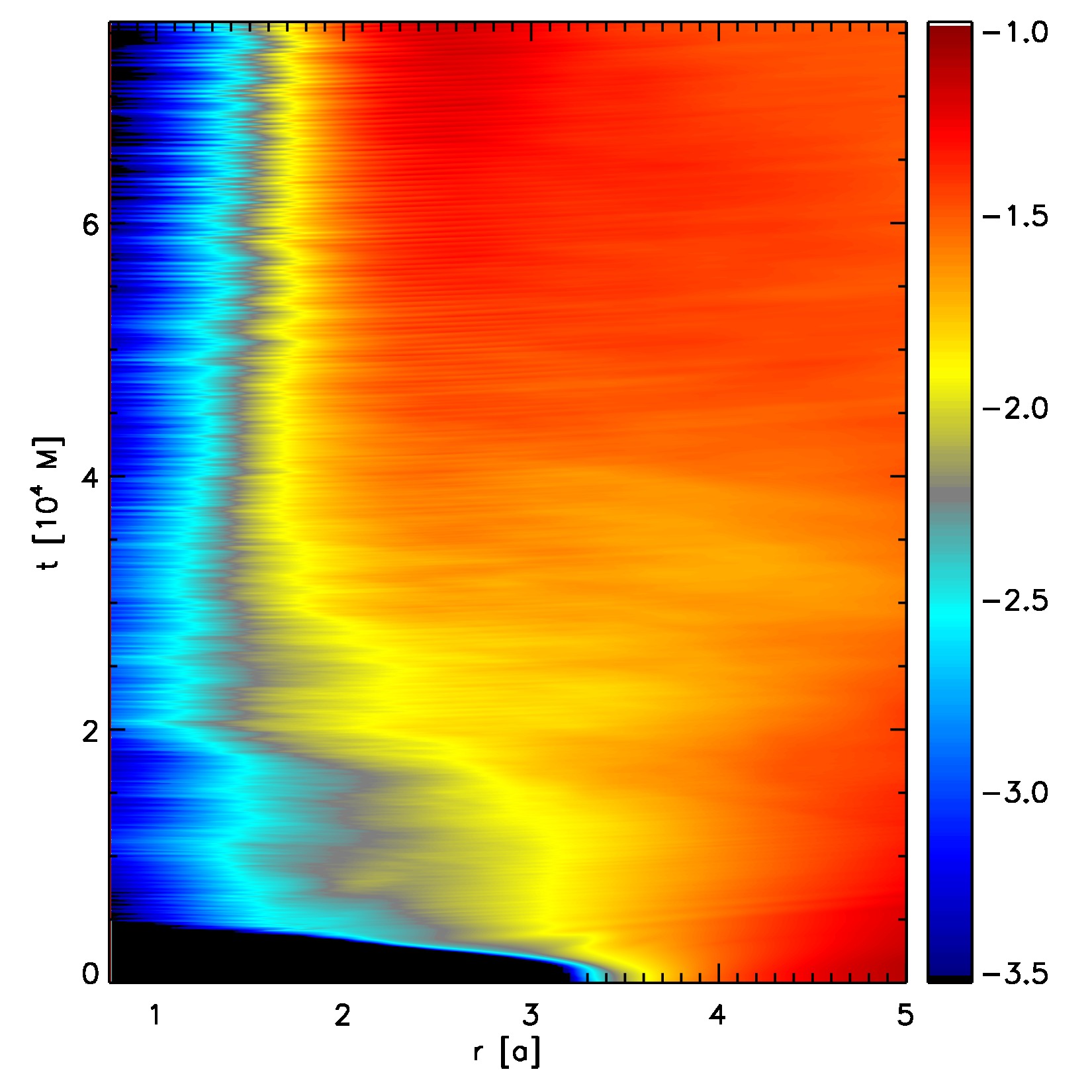}
  \caption[]{Color contour of surface density in logarithmic units, $\log_{10}
    \Sigma(t,r)$. Left panel: \mhdonepnnew evolution; right panel: \mhdtwopn
    evolution. \label{fig:mhd-surf-dens_tr} }
\end{figure*}

\begin{figure*}[tbhp]
  \centering
  \includegraphics[width=0.32\textwidth]{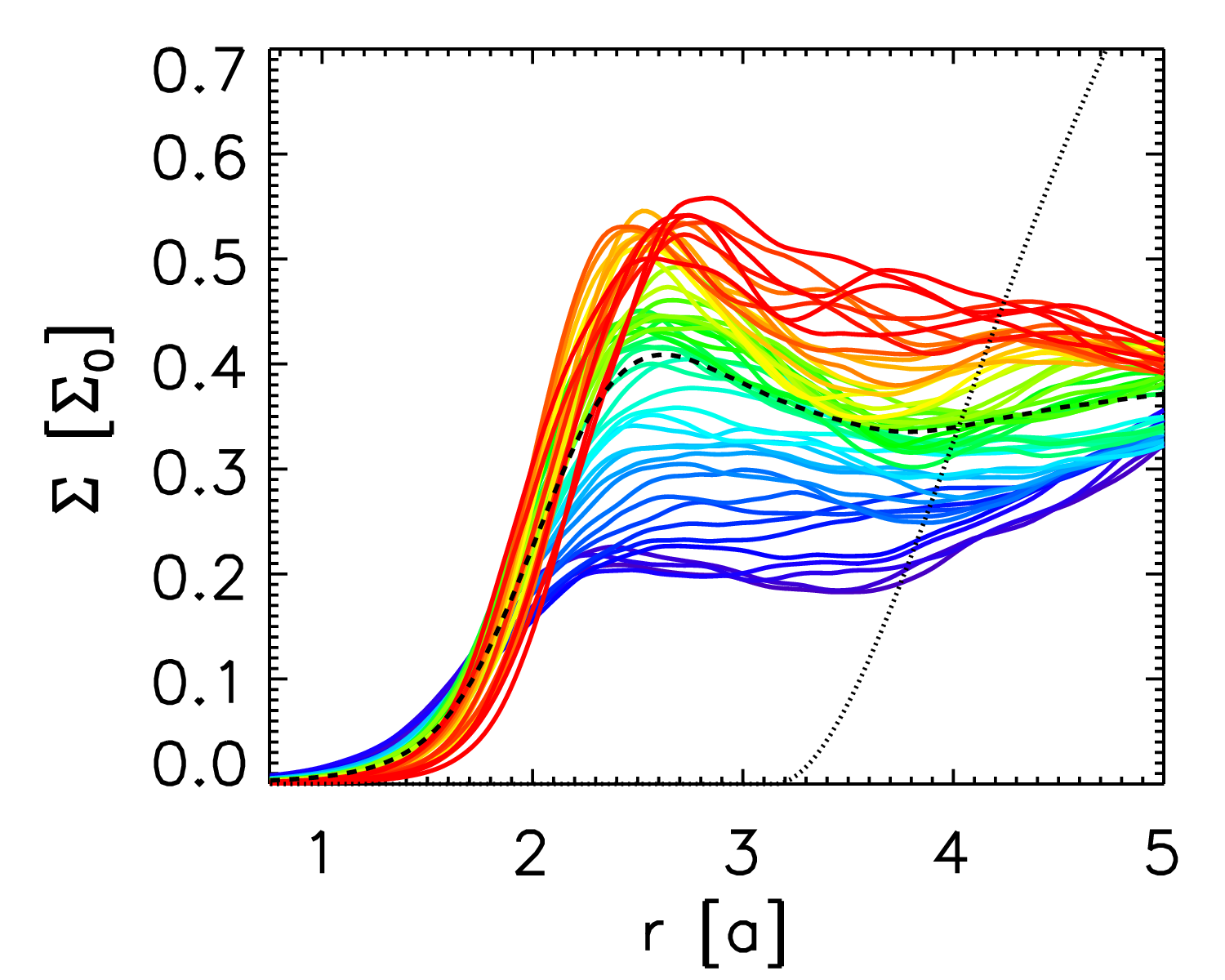}
  \includegraphics[width=0.32\textwidth]{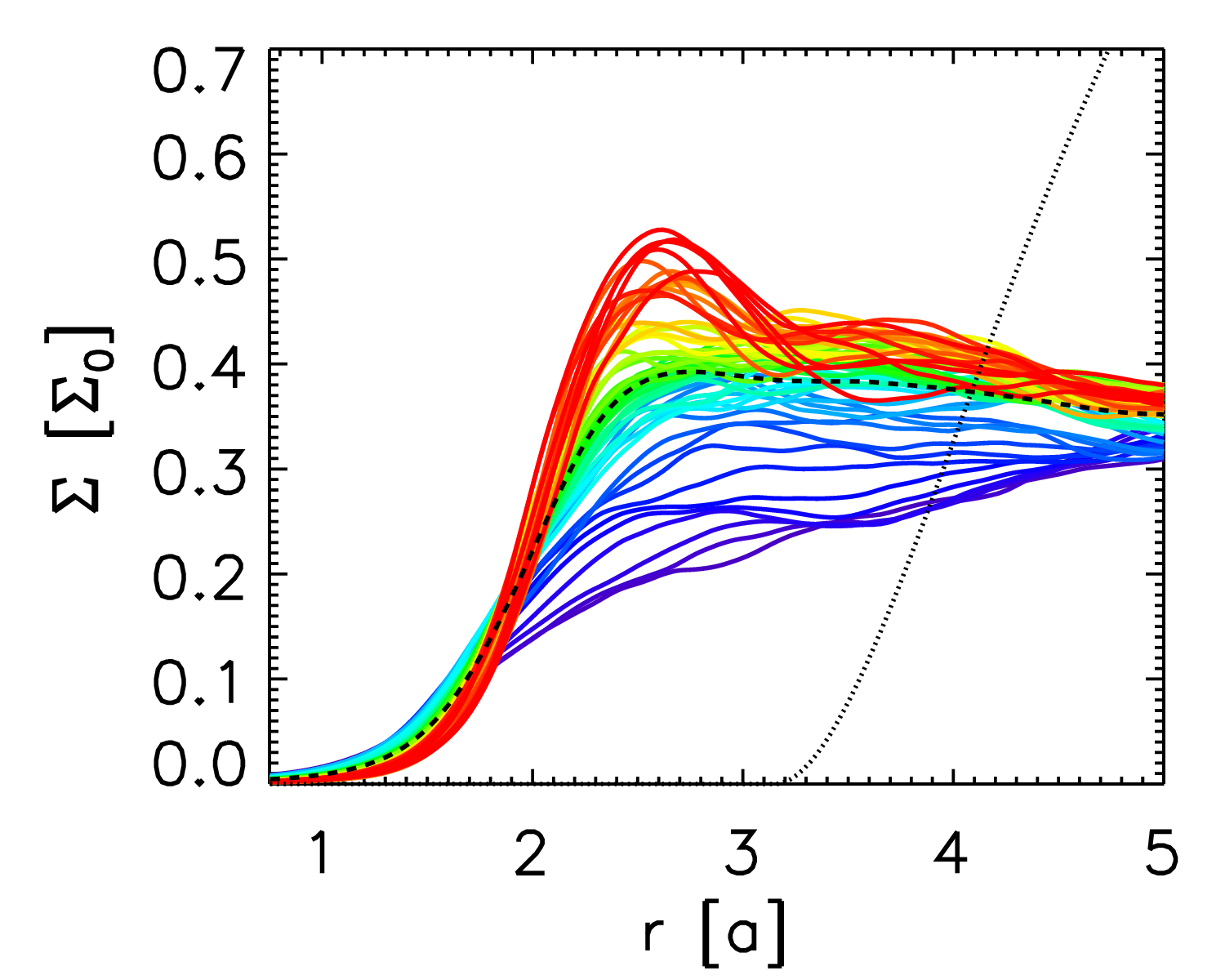}
  \includegraphics[width=0.32\textwidth]{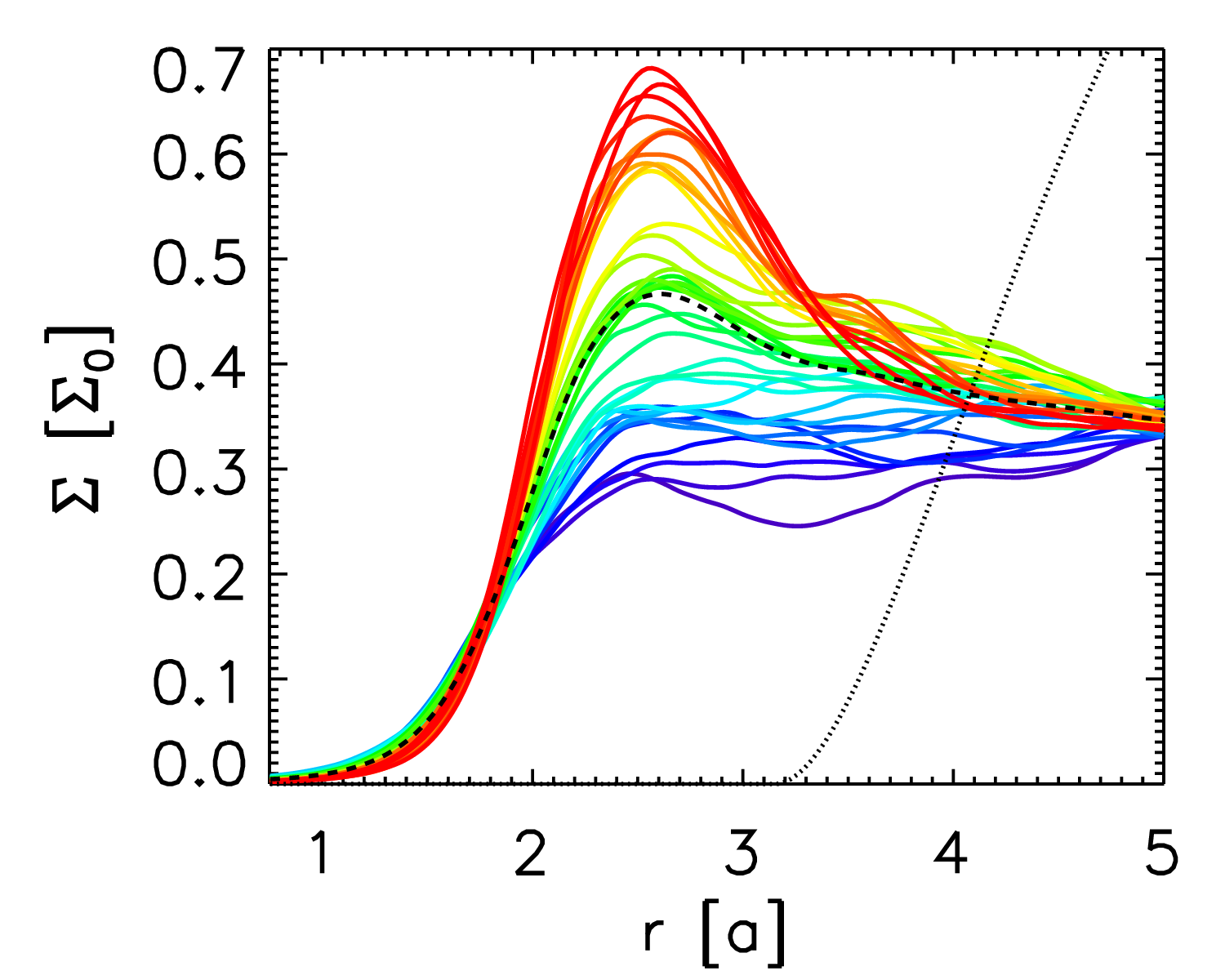}
  \caption[]{Surface density $\Sigma(r/a)$ for fixed time steps, from beginning
    of the secularly evolving state (blue lines, bottom) to end of the simulation (red lines, top). The dotted curve shows
    the initial condition, while the dashed curved shows the average of the
    colored curves. Left panel: \mhdonepnold; middle panel: \mhdonepnnew; right
    panel: \mhdtwopn.
    \label{fig:mhd-surf-dens} }
\end{figure*}

Figure~\ref{fig:mhd-surf-dens_tr} shows color contour plots of the surface density in logarithmic units, $\log_{10} \Sigma(t,r)$, for both \mhdonepnnew and \mhdtwopn evolutions.
At around $r \simeq 2.5a$ we observe a steady increase in the surface density with time that eventually plateaus. 
Unlike the cases of Sec.~\ref{sec:surf-dens}, here we see there are significant departures from initial conditions as we would expect because of the magnetic field
providing a means of efficient angular momentum transfer.  On the logarithmic scale, however, we see now only minor differences between the two scenarios. 
When analyzing Fig.~\ref{fig:mhd-surf-dens} (which shows line plots of the surface density as a function of radial distance for fixed time steps), however, it seems that a distinctive pattern forms in the \mhdtwopn case, where a distinct local maximum appears at $r\simeq 2.5a$ after a certain time, and persisting until the end of the evolution.
Such a pattern is not as pronounced in the \mhdonepnnew case. 
We note, however, that in the \mhdonepnold evolution (which has the same angular momentum as the \mhdtwopn one), this same pattern is indeed observed, implying
that this difference is likely due to the differences in the disks' initial angular momenta.

\subsection{Mass enclosed}
\label{sec:mhd-mass-en}

\begin{figure*}[tbhp]
  \centering
  \includegraphics[width=0.45\textwidth]{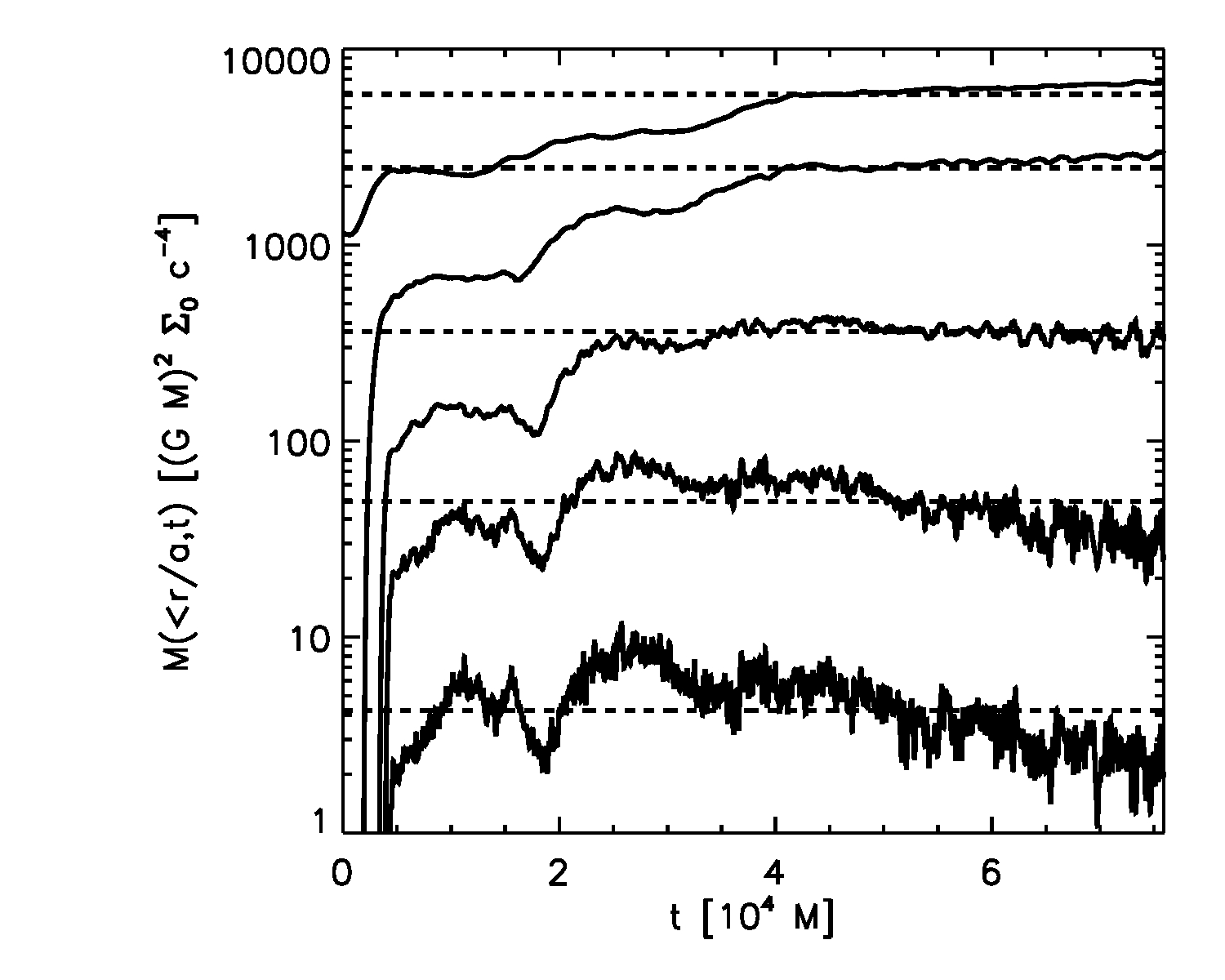}
  \includegraphics[width=0.45\textwidth]{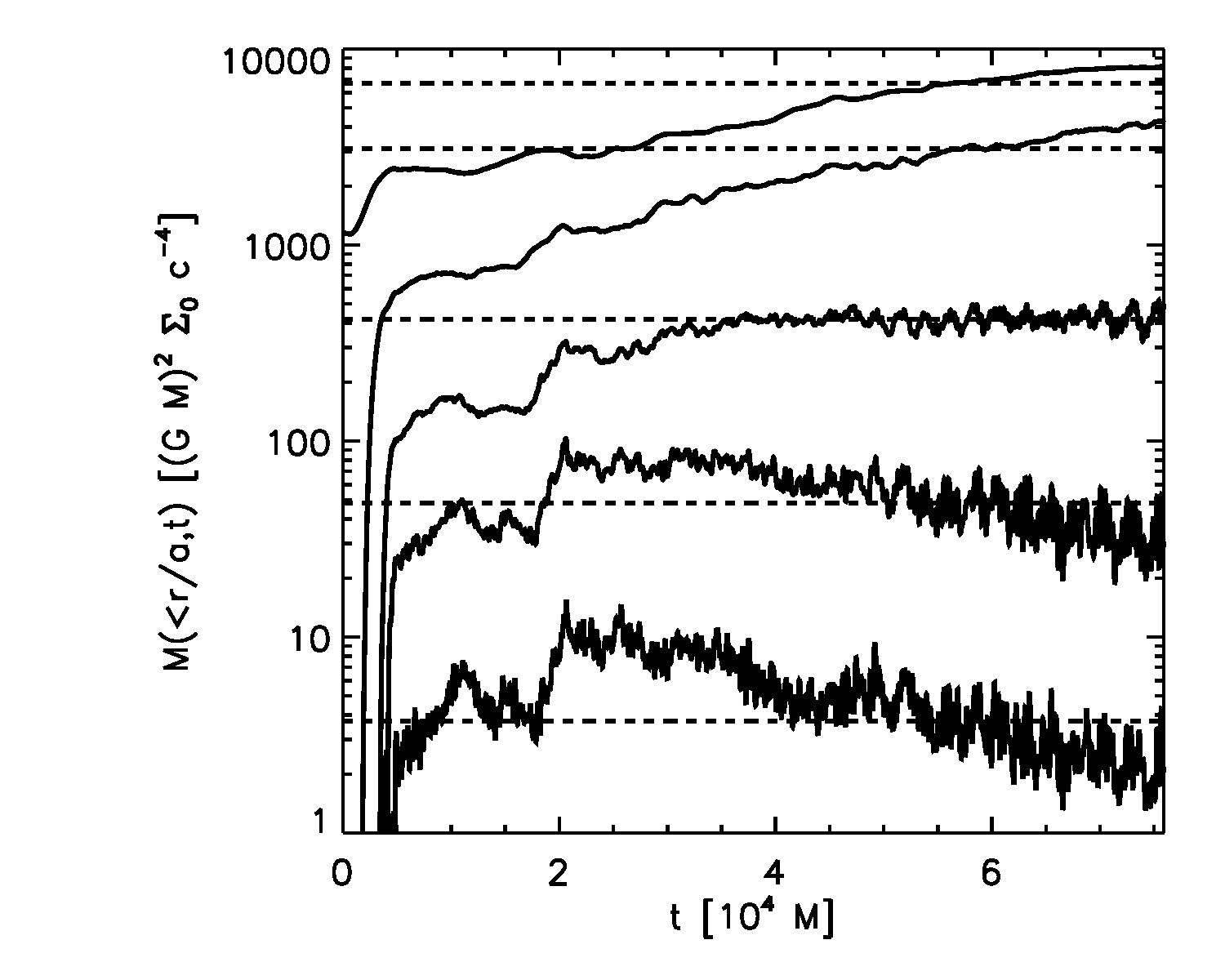}
  \caption[]{Mass enclosed within $r<(1,1.5,2,3,4)a$ (bottom to top,
    respectively) in logarithmic units for simulations \mhdonepnnew (left panel)
    and \mhdtwopn (right panel). \label{fig:mhd-mass-enc} }
\end{figure*}

In Fig.~\ref{fig:mhd-mass-enc} we have the equivalent of Fig.~\ref{fig:mass-enc-a50-20} for our MHD evolutions, where we see how matter accumulates in the inner disk over time.
For both situations, we see a sharp increase in the mass enclosed at smaller radii early on in the evolution. As the initial transient fades away and the disk evolves to a quasi-equilibrium configuration, the mass stabilizes at a roughly constant value, particularly for large radii.

As was the case for the previous quantities, we see no noticeable difference between 1PN and 2.5PN cases.  We therefore find that the differences in the 
spacetime and the disks' initial conditions to not have a large effect on the rates of achieving a degree of mass inflow equilibrium.

\subsection{Luminosity}
\label{sec:luminosity}

As in~\cite{Noble:2012xz}, we employ a local radiative cooling function to 
control a disk's aspect ratio and provide a means to predict bolometric electromagnetic signatures of 
circumbinary flows for a specified disk scale height profile.  Specifically, the 
cooling function is designed so that the gas loses heat whenever its entropy rises above the initial 
constant entropy of the disk.  This way, the integrated luminosity over time is a record of the total 
energy dissipated by the gas.  
Each disk was initialized to have approximately 
the same scale height.  Specifically, \mhdonepnnew and \mhdtwopn started with a aspect ratio of $\simeq 0.1$ within an accuracy less than 1\%, 
while \mhdonepnold started with a slightly slenderer profile, $\simeq 0.09$.   In order to uncover any  photometric 
predictors for a circumbinary system, we have analyzed the light curves produced by the simulations.  We estimate 
the total luminosity of the disk from the local emissivity (cooling function) via 
\begin{equation}
  \label{eq:cooling}
  L(t) = \int \sqrt{-g} d\theta d\phi dr \mathcal{L}_c u_t \, . 
\end{equation}
This method is approximate because it assumes that the radiation reaches the observer immediately without any 
relativistic redshift or delay (e.g., because of the disk's opacity).  It remains to be seen if these assumptions are significant, but  
there are many reasons to trust our results. 
For instance, the largest relativistic 
effects are expected to  occur immediately near the black holes---a region we excise from our simulations; our results are therefore useful
to investigate variability from the circumbinary disk and not from the matter within the gap.  Further, including opacity is expected
to weaken the quality factor of any periodic signal we observe as it will introduce an incoherent delay, so minor differences in the light 
curves will likely be obfuscated by this effect. 

\begin{figure*}[tbhp]
  \centering
  \includegraphics[width=0.32\textwidth]{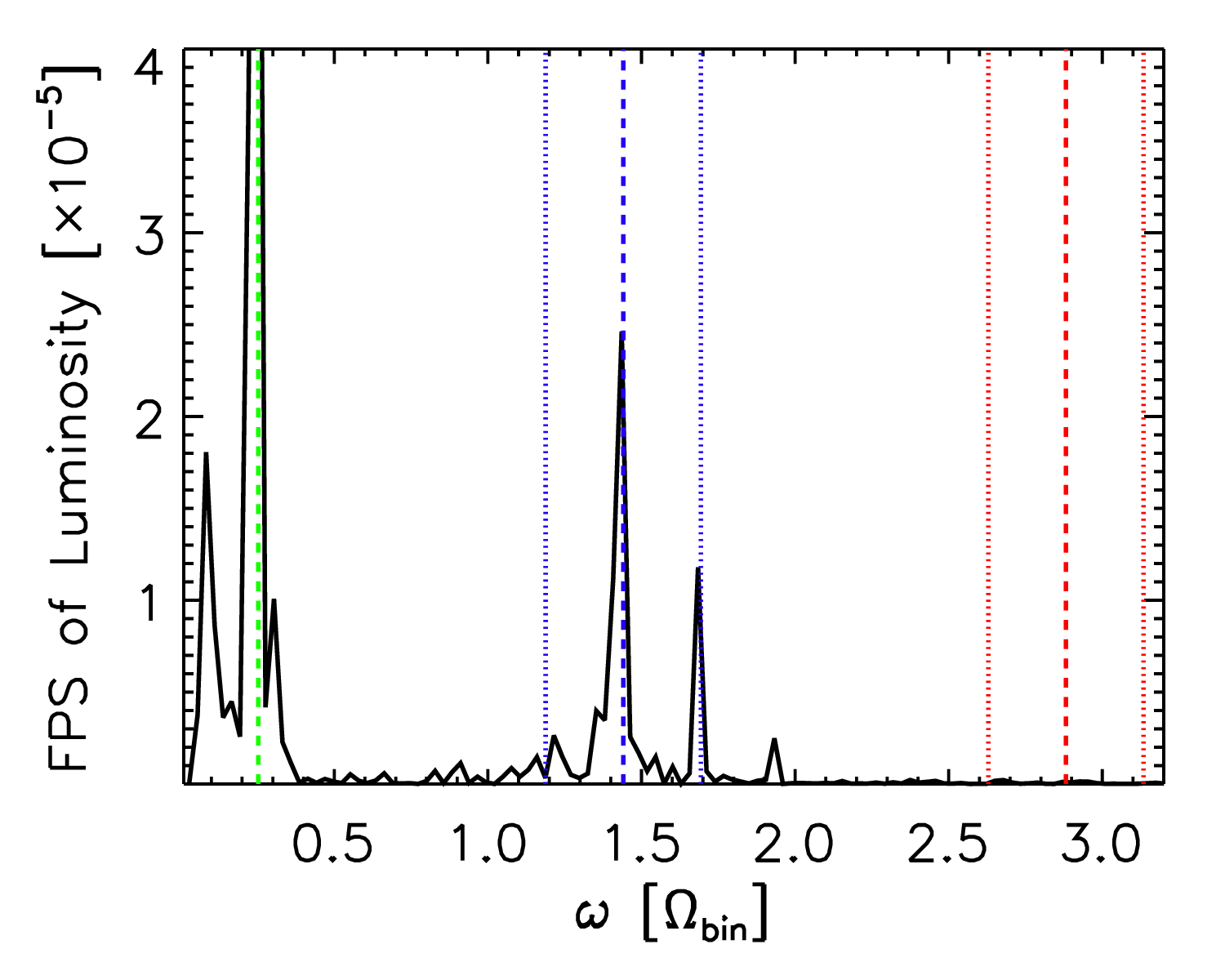}
  \includegraphics[width=0.32\textwidth]{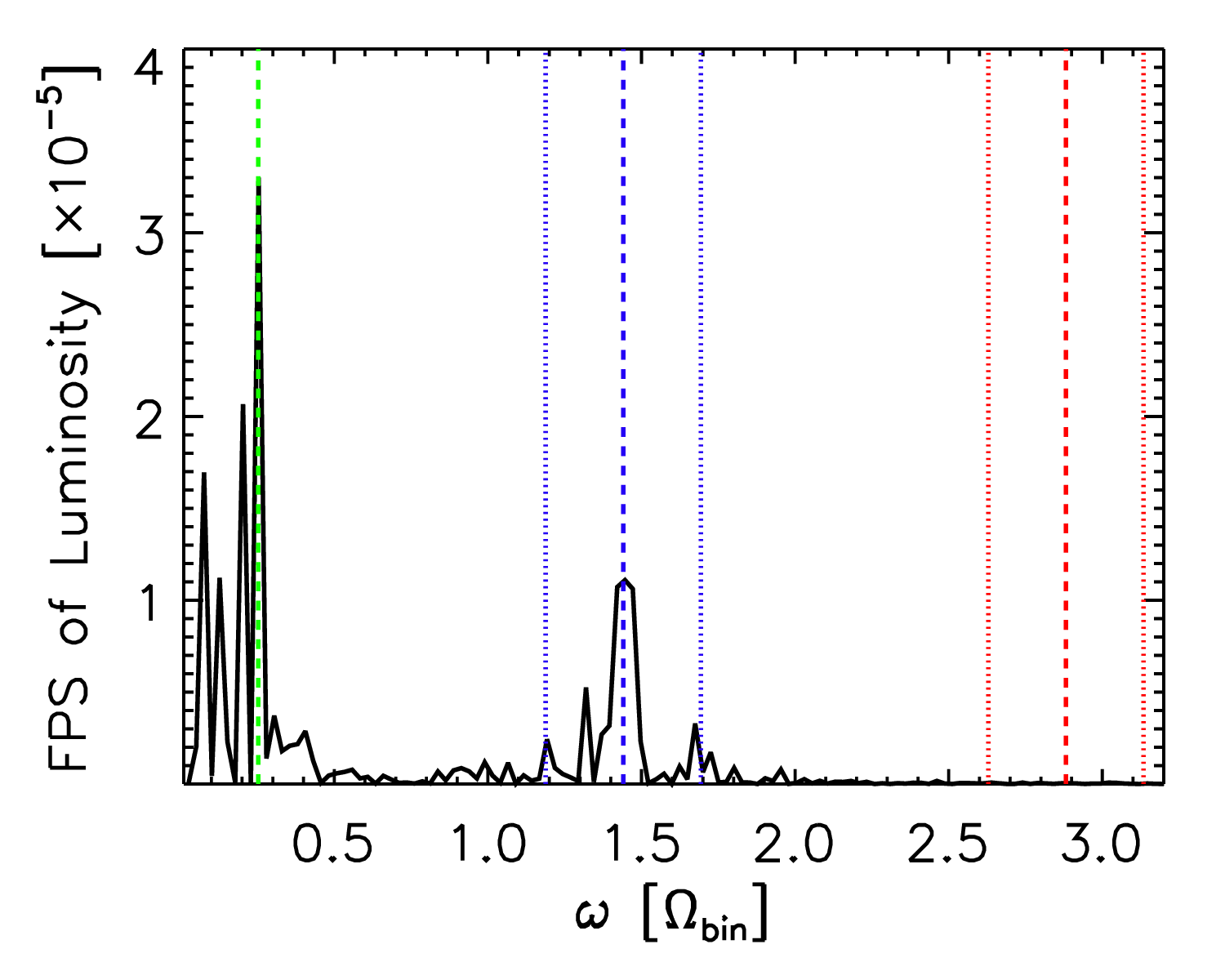}
  \includegraphics[width=0.32\textwidth]{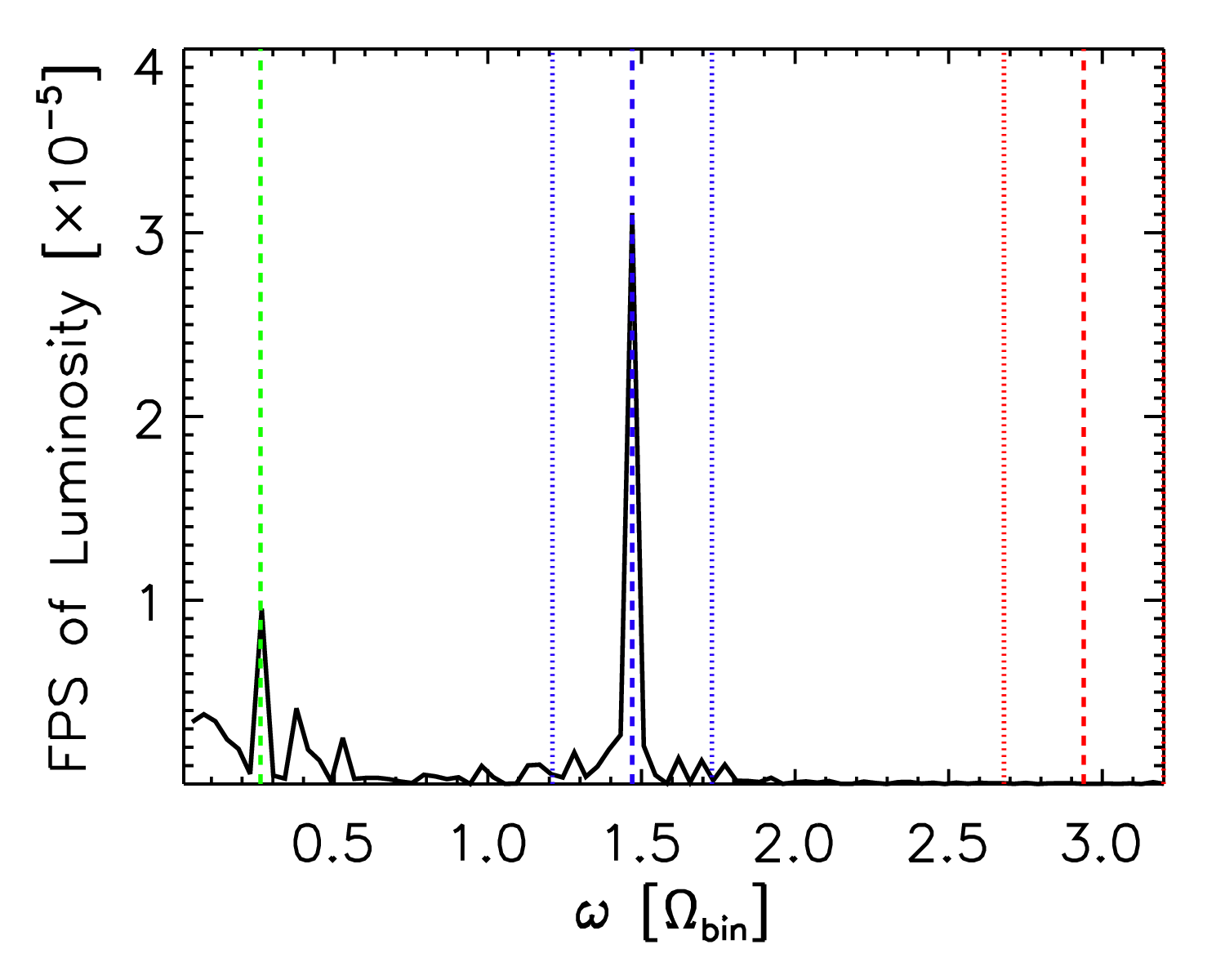}
  \caption[]{FFT of luminosity for our MHD simulations \mhdonepnold (left), \mhdonepnnew (middle), \mhdtwopn (right).
    Highlighted frequencies include  the 
    orbital frequency at time-average radius of $\max(\Sigma) \equiv \Omega_1$ (green line); 
    $\Omega_2 \equiv 1.375 \Omega\_{\rm bin}$ (blue dashes);     
    $\Omega_1 \pm \Omega_2$ (blue dots); 
    and the overtones of $\Omega_1 \pm \Omega_2$ (red dots) and $\Omega_2$ (red dashes). Green, blue and red lines appear in the
    leftmost, central and rightmost part of the figure respectively.
    \label{fig:mhd-luminosity} }
\end{figure*}

When comparing any two simulations involving chaotic turbulent flows, one never expects them to be 
able to make quantitative timestep-to-timestep comparisons if they start with small differences.  We therefore present, 
temporal power spectrum (Fast Fourier Transform, FFT, of $L(t)$) from each simulation's light curve instead of the light curves themselves.  
In Fig.~\ref{fig:mhd-luminosity}  we plot the Fourier power spectrum of $L(t)$ over the latter part of 
the secularly-evolving period, from $t\simeq60,000M$ onward.  We find that each simulation exhibits the 
same strong periodic signal found well above the background noise of fluctuations.  
As we identified before in \cite{Noble:2012xz}, it occurs at twice the beat frequency between the binary's orbit and the 
orbit of the non-axisymmetric overdensity feature that develops at $r\simeq 2.5a$. The frequency, $\Omega_1$, appears at 
$\Omega_1 \simeq 1.47 \Omega_\mathrm{bin}$ for the \mhdtwopn run, and  $\Omega_1 \simeq 1.44 \Omega_\mathrm{bin}$ for the 
two 1PN runs.  Variability is seen at the frequency of the
overdensity's orbit, which we call $\Omega_2$. 
The time-averaged radial coordinate of the overdensity $r_\mathrm{max}/a \simeq 2.56$ for \mhdtwopn and $r_\mathrm{max} \simeq 2.61$ for the 
1PN runs, leading to $\Omega_2 \simeq 0.26 \Omega_\mathrm{bin}$ for \mhdtwopn and $\Omega_2 \simeq 0.25 \Omega_\mathrm{bin}$ for the 
1PN runs. 

All three simulations see the signal at $\Omega_1$, though the \mhdtwopn run exhibits the clearest peak.  On the other hand, the 1PN runs exhibit more power 
at $\Omega_2$ and $\Omega_1 \pm \Omega_2$, suggesting that the variability in these runs stems more from the orbital motion of the overdensity and not the 
coherent interaction between the overdensity and the binary.  Both 1PN runs demonstrate more power at the overdensity orbital frequency than at any other 
frequency, unlike what is seen from \mhdtwopn.  This enhancement at $\Omega_2$ seen in the 1PN runs is unlikely due to greater low-frequency 
noise because the 1PN power spectra  were calculated from longer periods of $L(t)$ data than what was used for \mhdtwopn's spectrum. Even though more 
runs would be needed for a definitive answer, the disparity in variability power at $\Omega_1$ and $\Omega_2$ is the most consistent difference seen 
in the light curves from 1PN and 2.5PN simulations.

\section{Final remarks}
\label{sec:final}

In this work, we explored how PN order affects the evolution of 
non-magnetized and magnetized gas around binary black hole systems.
For inviscid hydrodynamics, and separations  $a\gtrsim 50M$, we  found
only very small differences in the gas dynamics between 1PN and 2.5PN
spacetimes.  For smaller separations ($a\lesssim 40M$), there are
noticeable differences in the transient behavior of the disk between PN
orders, but even down to $a\simeq 30M$ there are very little
differences in the bulk dynamics of the disk. 
At separations of $a\lesssim 20M$, there are noticeable differences in
the bulk dynamics of unmagnetized gas between the 1PN and 2.5PN
spacetimes. 

We next looked at full MHD simulations of magnetized
disks at $a=20M$. We performed two MHD simulations using the 1PN
spacetime to compare with the results obtained from 2.5PN accurate
spacetime, published in~\cite{Noble:2012xz}. We only
observed small differences between all three MHD simulations in
the bulk of the disk,
e.g. the differences between the 1PN and 2.5PN cases are of the same
order of that between the two 1PN runs. This leads us to conclude that
differences between 1PN and 2.5PN are of the same order of magnitude
as differences that one would find from different initial conditions.
This is because the MHD dynamics, which drives accretion, 
seems to effectively mask the effects from the high-order PN terms.  In all three
MHD runs, we discovered a unique and exciting periodic EM signature
that could be used to both identify SMBH mergers in the time domain
and measure their mass ratio.  This signal is robust down to small
binary separations, such as 20M, though it is the strongest 
signal over the entire frequency range for only 
the 2.5PN order simulation.  
Of course, it remains to be seen if
the quantitative differences are larger than the systematic error arising from
our choice of initial conditions and our choice to excise the binary. This will require further
studies and simulations that are beyond the scope of this paper.


While the bulk of the disk is largely unaffected by PN order, the
surface density at the inner edge of the disk shows a more significant
lump for 2.5PN than 1PN. These differences  are most likely due to
enhanced torque densities in the
2.5PN metric within the gap.  If one is interested in understanding
the physics at the interface between the gap and the inner edge of the
disk, our results suggest that the 2.5PN metric should be used at
separations of 20M and smaller.
This result, is particularly interesting in
the context of a new type of simulation we are exploring, where 
each BH resides on the
numerical domain. With this new study, we intend to explore how 
mini-disks form, how the accreting matter is distributed about the two
SMBHs, and how the orbital dynamics of the BHs is affected by accretion. The
distribution of gas and dissipation of internal stresses will provide
us with the means of tracking when and where light is radiated in the
system and answer key questions about the accretion dynamics of
merging SMBHs.

\begin{acknowledgments}
  We would like to thank J.~Krolik, B.~Mundim, H.~Nakano and L.~Blanchet for
  discussions and helpful input.
  M.Z. is supported by NSF grants OCI-0832606, PHY-0969855, AST-1028087, and
  PHY-1229173.  S.C.N. is supported by NSF grant No. OCI-0725070, OCI-0832606,
  AST-1028087, PHY-1125915.  Computational resources were provided by XSEDE
  allocation TG-PHY060027N, and by NewHorizons and BlueSky Clusters at Rochester
  Institute of Technology, which were supported by NSF grant No. PHY-0722703,
  DMS-0820923, AST-1028087, and PHY-1229173.
\end{acknowledgments}

\appendix

\section{Numerical Details}
\label{app:details}

\begin{figure*}[tbhp]
  \centering
  \includegraphics[width=0.45\textwidth]{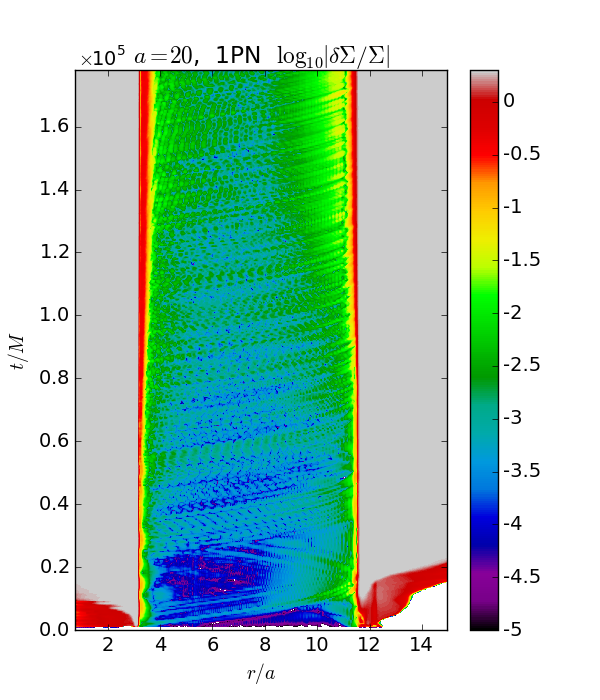} 
  \includegraphics[width=0.45\textwidth]{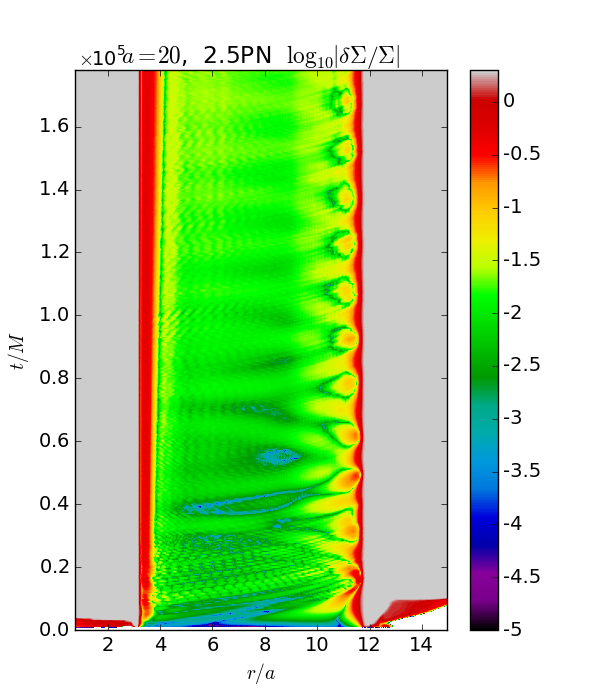}
  \caption[]{
    High resolution version of lower panel of Fig.~\ref{fig:delta_sigma_vs_r_t-a30-a20}. We
    repeated runs \hydroonepn{20} and \hydrotwopn{20} using $480 \times 480$ cells, in order to
    test the accuracy of our evolution.
    \label{fig:delta_sigma_vs_r_t-a20_hres}
  }
\end{figure*}

In order to make sure that our results were not an artifact of numerical errors accumulating with time, we have repeated runs \hydroonepn{20} and \hydrotwopn{20} using $480 \times 480$ cells (instead of our ``standard'' $320 \times 320$). In Fig.~\ref{fig:delta_sigma_vs_r_t-a20_hres} we plot the corresponding relative change in $\Sigma$, since this quantity proved to be quite sensitive to small changes in configurations. These plots should be matched against the lower panel of Fig.~\ref{fig:delta_sigma_vs_r_t-a30-a20}, depicting its lower resolution counterpart. As can be observed, the figures are remarkably similar, which gives us confidence that numerical errors are not masking the results we have found.

Another approximation we make in our non-magnetized and magnetized runs is that we 
excise the spherical region including the binary's orbit, out to a \emph{coordinate}
radius of 1.5 binary separations.  One may argue that differences in the two PN spacetimes
will put this excision surface at different proper distances, and so may result in an 
inconsistent setup that ultimately contaminates the numerical 
comparisons between runs with different PN accuracies.  As one can see from 
the relative differences in $g_{rr}$ shown in Figure~\ref{fig:a20_a100_t0}, the 
differences in proper distances beyond the binary's orbit are expected
to be no more than a few percent. Further,~\cite{Shi:2011us} have performed
a sequence of Newtonian MHD runs with different excision radii and found that changes $>10\%$ 
yield insignificant differences in results, such as those reported here.  Further, their 
Newtonian results---specifically the various contributions to $d^2J/dt dr$---are remarkably
similar to our 2.5PN results, let alone our 1PN results.  Hence, we do not expect that 
differences in the proper radius of the excision surface to be a large effect.  Whether or 
not the excision method is valid at all requires us to repeat our simulations without any 
excision.  Such calculations would be extremely expensive to perform in their entirety, but 
we hope to investigate this issue with shorter runs and/or smaller setups in the future. 

The principal angular momentum transfer mechanism is the correlated MHD stress associated with the MRI.  It is therefore critical 
for simulations hoping to realistically represent a magnetized turbulent disk to adequately resolve the fastest growing modes of 
this instability.   Experience has shown that several conditions should be met in order for global simulations to sufficiently resolve 
the MRI \cite{Noble:2010mm,hgk11} and the vertical scale height of the disk \cite{2012ApJ...749..189S}.   The MRI resolution 
benchmarks are given in terms of quality factors per dimension ($Q^{(i)}$) 
which are proportional to the number of cells per MRI wavelength along 
the dimension.  Their suggested benchmarks are $Q^{(\theta)}  > 10$ and  $Q^{(\phi)} > 25$ \cite{hgk11}, which all our MHD 
simulations meet everywhere in the disk's bulk until late times ($t \gtrsim 60,000M$) in the orbiting overdensity region or ``lump.'' 
Further, all our simulations cover the disk's  vertical scale height with more than $36$ cells on average 
over the course of each run, which more than satisfies the suggested target of $32$ cells per scale height of \cite{2012ApJ...749..189S}. 
We emphasize that our new results presented here satisfy these resolution requirements just as well as our original run, hence 
all our MHD results resolve the MRI quite well.  Please see \cite{Noble:2012xz} for specific details on how we 
measure a run's MRI resolution criteria.   Again, we wish to emphasize 
that the only uncertainty in our simulations meeting the MRI resolution criteria arises at late times when magnetic flux escapes and/or
dissipates within the overdensity region;  at all other times throughout the bulk of the disk, the criteria are satisfied 
with great excess.  We will explore  flux loss phenomenon in the lump using a different variety of initial conditions 
in a future paper.


\bibliography{references}

\begin{thebibliography}{43}
\expandafter\ifx\csname natexlab\endcsname\relax\def\natexlab#1{#1}\fi
\expandafter\ifx\csname bibnamefont\endcsname\relax
  \def\bibnamefont#1{#1}\fi
\expandafter\ifx\csname bibfnamefont\endcsname\relax
  \def\bibfnamefont#1{#1}\fi
\expandafter\ifx\csname citenamefont\endcsname\relax
  \def\citenamefont#1{#1}\fi
\expandafter\ifx\csname url\endcsname\relax
  \def\url#1{\texttt{#1}}\fi
\expandafter\ifx\csname urlprefix\endcsname\relax\def\urlprefix{URL }\fi
\providecommand{\bibinfo}[2]{#2}
\providecommand{\eprint}[2][]{\url{#2}}

\bibitem[{\citenamefont{{Magorrian} et~al.}(1998)\citenamefont{{Magorrian},
  {Tremaine}, {Richstone}, {Bender}, {Bower}, {Dressler}, {Faber}, {Gebhardt},
  {Green}, {Grillmair} et~al.}}]{Magorrian1998}
\bibinfo{author}{\bibfnamefont{J.}~\bibnamefont{{Magorrian}}},
  \bibinfo{author}{\bibfnamefont{S.}~\bibnamefont{{Tremaine}}},
  \bibinfo{author}{\bibfnamefont{D.}~\bibnamefont{{Richstone}}},
  \bibinfo{author}{\bibfnamefont{R.}~\bibnamefont{{Bender}}},
  \bibinfo{author}{\bibfnamefont{G.}~\bibnamefont{{Bower}}},
  \bibinfo{author}{\bibfnamefont{A.}~\bibnamefont{{Dressler}}},
  \bibinfo{author}{\bibfnamefont{S.~M.} \bibnamefont{{Faber}}},
  \bibinfo{author}{\bibfnamefont{K.}~\bibnamefont{{Gebhardt}}},
  \bibinfo{author}{\bibfnamefont{R.}~\bibnamefont{{Green}}},
  \bibinfo{author}{\bibfnamefont{C.}~\bibnamefont{{Grillmair}}},
  \bibnamefont{et~al.}, \bibinfo{journal}{\aj} \textbf{\bibinfo{volume}{115}},
  \bibinfo{pages}{2285} (\bibinfo{year}{1998}), \eprint{astro-ph/9708072}.

\bibitem[{\citenamefont{{Gebhardt} et~al.}(2000)\citenamefont{{Gebhardt},
  {Bender}, {Bower}, {Dressler}, {Faber}, {Filippenko}, {Green}, {Grillmair},
  {Ho}, {Kormendy} et~al.}}]{Gebhardt2000}
\bibinfo{author}{\bibfnamefont{K.}~\bibnamefont{{Gebhardt}}},
  \bibinfo{author}{\bibfnamefont{R.}~\bibnamefont{{Bender}}},
  \bibinfo{author}{\bibfnamefont{G.}~\bibnamefont{{Bower}}},
  \bibinfo{author}{\bibfnamefont{A.}~\bibnamefont{{Dressler}}},
  \bibinfo{author}{\bibfnamefont{S.~M.} \bibnamefont{{Faber}}},
  \bibinfo{author}{\bibfnamefont{A.~V.} \bibnamefont{{Filippenko}}},
  \bibinfo{author}{\bibfnamefont{R.}~\bibnamefont{{Green}}},
  \bibinfo{author}{\bibfnamefont{C.}~\bibnamefont{{Grillmair}}},
  \bibinfo{author}{\bibfnamefont{L.~C.} \bibnamefont{{Ho}}},
  \bibinfo{author}{\bibfnamefont{J.}~\bibnamefont{{Kormendy}}},
  \bibnamefont{et~al.}, \bibinfo{journal}{\apjl}
  \textbf{\bibinfo{volume}{539}}, \bibinfo{pages}{L13} (\bibinfo{year}{2000}),
  \eprint{astro-ph/0006289}.

\bibitem[{\citenamefont{{Ferrarese} and
  {Merritt}}(2000)}]{FerrareseMerritt2000}
\bibinfo{author}{\bibfnamefont{L.}~\bibnamefont{{Ferrarese}}} \bibnamefont{and}
  \bibinfo{author}{\bibfnamefont{D.}~\bibnamefont{{Merritt}}},
  \bibinfo{journal}{\apjl} \textbf{\bibinfo{volume}{539}}, \bibinfo{pages}{L9}
  (\bibinfo{year}{2000}), \eprint{astro-ph/0006053}.

\bibitem[{\citenamefont{{Kormendy} and {Ho}}(2013)}]{KormendyHo2013}
\bibinfo{author}{\bibfnamefont{J.}~\bibnamefont{{Kormendy}}} \bibnamefont{and}
  \bibinfo{author}{\bibfnamefont{L.~C.} \bibnamefont{{Ho}}},
  \bibinfo{journal}{Ann. Rev. Astron. Astrop.} \textbf{\bibinfo{volume}{51}},
  \bibinfo{pages}{511} (\bibinfo{year}{2013}), \eprint{1304.7762}.

\bibitem[{\citenamefont{{Silk} and {Rees}}(1998)}]{SilkRees1998}
\bibinfo{author}{\bibfnamefont{J.}~\bibnamefont{{Silk}}} \bibnamefont{and}
  \bibinfo{author}{\bibfnamefont{M.~J.} \bibnamefont{{Rees}}},
  \bibinfo{journal}{\aap} \textbf{\bibinfo{volume}{331}}, \bibinfo{pages}{L1}
  (\bibinfo{year}{1998}), \eprint{astro-ph/9801013}.

\bibitem[{\citenamefont{{Fabian}}(1999)}]{Fabian1999}
\bibinfo{author}{\bibfnamefont{A.~C.} \bibnamefont{{Fabian}}},
  \bibinfo{journal}{\mnras} \textbf{\bibinfo{volume}{308}},
  \bibinfo{pages}{L39} (\bibinfo{year}{1999}), \eprint{astro-ph/9908064}.

\bibitem[{\citenamefont{{Volonteri} et~al.}(2003)\citenamefont{{Volonteri},
  {Haardt}, and {Madau}}}]{Vol03}
\bibinfo{author}{\bibfnamefont{M.}~\bibnamefont{{Volonteri}}},
  \bibinfo{author}{\bibfnamefont{F.}~\bibnamefont{{Haardt}}}, \bibnamefont{and}
  \bibinfo{author}{\bibfnamefont{P.}~\bibnamefont{{Madau}}},
  \bibinfo{journal}{\apj} \textbf{\bibinfo{volume}{582}}, \bibinfo{pages}{559}
  (\bibinfo{year}{2003}), \eprint{arXiv:astro-ph/0207276}.

\bibitem[{\citenamefont{{Volonteri}}(2007)}]{Vol07}
\bibinfo{author}{\bibfnamefont{M.}~\bibnamefont{{Volonteri}}},
  \bibinfo{journal}{\apjl} \textbf{\bibinfo{volume}{663}}, \bibinfo{pages}{L5}
  (\bibinfo{year}{2007}), \eprint{arXiv:astro-ph/0703180}.

\bibitem[{\citenamefont{{Schnittman}}(2007)}]{Schnitt07}
\bibinfo{author}{\bibfnamefont{J.~D.} \bibnamefont{{Schnittman}}},
  \bibinfo{journal}{\apjl} \textbf{\bibinfo{volume}{667}},
  \bibinfo{pages}{L133} (\bibinfo{year}{2007}), \eprint{0706.1548}.

\bibitem[{\citenamefont{{Begelman} et~al.}(1980)\citenamefont{{Begelman},
  {Blandford}, and {Rees}}}]{BBR80}
\bibinfo{author}{\bibfnamefont{M.~C.} \bibnamefont{{Begelman}}},
  \bibinfo{author}{\bibfnamefont{R.~D.} \bibnamefont{{Blandford}}},
  \bibnamefont{and} \bibinfo{author}{\bibfnamefont{M.~J.}
  \bibnamefont{{Rees}}}, \bibinfo{journal}{\nat}
  \textbf{\bibinfo{volume}{287}}, \bibinfo{pages}{307} (\bibinfo{year}{1980}).

\bibitem[{\citenamefont{Deane et~al.}(2014)\citenamefont{Deane, Paragi, Jarvis,
  Coriat, Bernardi et~al.}}]{Deane:2014jqa}
\bibinfo{author}{\bibfnamefont{R.}~\bibnamefont{Deane}},
  \bibinfo{author}{\bibfnamefont{Z.}~\bibnamefont{Paragi}},
  \bibinfo{author}{\bibfnamefont{M.}~\bibnamefont{Jarvis}},
  \bibinfo{author}{\bibfnamefont{M.}~\bibnamefont{Coriat}},
  \bibinfo{author}{\bibfnamefont{G.}~\bibnamefont{Bernardi}},
  \bibnamefont{et~al.}, \bibinfo{journal}{Nature}
  \textbf{\bibinfo{volume}{511}}, \bibinfo{pages}{57 } (\bibinfo{year}{2014}),
  \eprint{1406.6365}.

\bibitem[{\citenamefont{Bogdanovic}(2014)}]{Bogdanovic:2014cua}
\bibinfo{author}{\bibfnamefont{T.}~\bibnamefont{Bogdanovic}}
  (\bibinfo{year}{2014}), \eprint{1406.5193}.

\bibitem[{\citenamefont{{Amaro-Seoane}
  et~al.}(2013)\citenamefont{{Amaro-Seoane}, {Aoudia}, {Babak}, {Bin{\'e}truy},
  {Berti}, {Boh{\'e}}, {Caprini}, {Colpi}, {Cornish}, {Danzmann}
  et~al.}}]{2013GWN.....6....4A}
\bibinfo{author}{\bibfnamefont{P.}~\bibnamefont{{Amaro-Seoane}}},
  \bibinfo{author}{\bibfnamefont{S.}~\bibnamefont{{Aoudia}}},
  \bibinfo{author}{\bibfnamefont{S.}~\bibnamefont{{Babak}}},
  \bibinfo{author}{\bibfnamefont{P.}~\bibnamefont{{Bin{\'e}truy}}},
  \bibinfo{author}{\bibfnamefont{E.}~\bibnamefont{{Berti}}},
  \bibinfo{author}{\bibfnamefont{A.}~\bibnamefont{{Boh{\'e}}}},
  \bibinfo{author}{\bibfnamefont{C.}~\bibnamefont{{Caprini}}},
  \bibinfo{author}{\bibfnamefont{M.}~\bibnamefont{{Colpi}}},
  \bibinfo{author}{\bibfnamefont{N.~J.} \bibnamefont{{Cornish}}},
  \bibinfo{author}{\bibfnamefont{K.}~\bibnamefont{{Danzmann}}},
  \bibnamefont{et~al.}, \bibinfo{journal}{GW Notes, Vol.~6, p.~4-110}
  \textbf{\bibinfo{volume}{6}}, \bibinfo{pages}{4} (\bibinfo{year}{2013}),
  \eprint{1201.3621}.

\bibitem[{\citenamefont{Amaro-Seoane et~al.}(2012)\citenamefont{Amaro-Seoane,
  Aoudia, Babak, Binetruy, Berti et~al.}}]{AmaroSeoane:2012je}
\bibinfo{author}{\bibfnamefont{P.}~\bibnamefont{Amaro-Seoane}},
  \bibinfo{author}{\bibfnamefont{S.}~\bibnamefont{Aoudia}},
  \bibinfo{author}{\bibfnamefont{S.}~\bibnamefont{Babak}},
  \bibinfo{author}{\bibfnamefont{P.}~\bibnamefont{Binetruy}},
  \bibinfo{author}{\bibfnamefont{E.}~\bibnamefont{Berti}},
  \bibnamefont{et~al.}, \bibinfo{journal}{Class. Quant. Grav.}
  \textbf{\bibinfo{volume}{29}}, \bibinfo{pages}{124016}
  (\bibinfo{year}{2012}), \eprint{1202.0839}.

\bibitem[{\citenamefont{Seoane et~al.}(2013)}]{Seoane:2013qna}
\bibinfo{author}{\bibfnamefont{P.~A.} \bibnamefont{Seoane}}
  \bibnamefont{et~al.} (\bibinfo{collaboration}{eLISA Collaboration})
  (\bibinfo{year}{2013}), \eprint{1305.5720}.

\bibitem[{\citenamefont{Macfadyen and Milosavljevic}(2008)}]{Macfadyen:2006jx}
\bibinfo{author}{\bibfnamefont{A.~I.} \bibnamefont{Macfadyen}}
  \bibnamefont{and}
  \bibinfo{author}{\bibfnamefont{M.}~\bibnamefont{Milosavljevic}},
  \bibinfo{journal}{Astrophys. J.} \textbf{\bibinfo{volume}{672}},
  \bibinfo{pages}{83} (\bibinfo{year}{2008}), \eprint{astro-ph/0607467}.

\bibitem[{\citenamefont{Shi et~al.}(2012)\citenamefont{Shi, Krolik, Lubow, and
  Hawley}}]{Shi:2011us}
\bibinfo{author}{\bibfnamefont{J.-M.} \bibnamefont{Shi}},
  \bibinfo{author}{\bibfnamefont{J.~H.} \bibnamefont{Krolik}},
  \bibinfo{author}{\bibfnamefont{S.~H.} \bibnamefont{Lubow}}, \bibnamefont{and}
  \bibinfo{author}{\bibfnamefont{J.~F.} \bibnamefont{Hawley}},
  \bibinfo{journal}{Astrophys. J.} \textbf{\bibinfo{volume}{749}},
  \bibinfo{pages}{118} (\bibinfo{year}{2012}), \eprint{1110.4866}.

\bibitem[{\citenamefont{Noble et~al.}(2012)\citenamefont{Noble, Mundim, Nakano,
  Krolik, Campanelli, Zlochower, and Yunes}}]{Noble:2012xz}
\bibinfo{author}{\bibfnamefont{S.~C.} \bibnamefont{Noble}},
  \bibinfo{author}{\bibfnamefont{B.~C.} \bibnamefont{Mundim}},
  \bibinfo{author}{\bibfnamefont{H.}~\bibnamefont{Nakano}},
  \bibinfo{author}{\bibfnamefont{J.~H.} \bibnamefont{Krolik}},
  \bibinfo{author}{\bibfnamefont{M.}~\bibnamefont{Campanelli}},
  \bibinfo{author}{\bibfnamefont{Y.}~\bibnamefont{Zlochower}},
  \bibnamefont{and} \bibinfo{author}{\bibfnamefont{N.}~\bibnamefont{Yunes}},
  \bibinfo{journal}{Astrophys. J.} \textbf{\bibinfo{volume}{755}},
  \bibinfo{pages}{51} (\bibinfo{year}{2012}), \eprint{1204.1073}.

\bibitem[{\citenamefont{{Roedig} and {Sesana}}(2014)}]{RoedigSesana14}
\bibinfo{author}{\bibfnamefont{C.}~\bibnamefont{{Roedig}}} \bibnamefont{and}
  \bibinfo{author}{\bibfnamefont{A.}~\bibnamefont{{Sesana}}},
  \bibinfo{journal}{\mnras} \textbf{\bibinfo{volume}{439}},
  \bibinfo{pages}{3476} (\bibinfo{year}{2014}), \eprint{1307.6283}.

\bibitem[{\citenamefont{{Farris} et~al.}(2014)\citenamefont{{Farris},
  {Duffell}, {MacFadyen}, and {Haiman}}}]{Farris14a}
\bibinfo{author}{\bibfnamefont{B.~D.} \bibnamefont{{Farris}}},
  \bibinfo{author}{\bibfnamefont{P.}~\bibnamefont{{Duffell}}},
  \bibinfo{author}{\bibfnamefont{A.~I.} \bibnamefont{{MacFadyen}}},
  \bibnamefont{and} \bibinfo{author}{\bibfnamefont{Z.}~\bibnamefont{{Haiman}}},
  \bibinfo{journal}{\apj} \textbf{\bibinfo{volume}{783}}, \bibinfo{eid}{134}
  (\bibinfo{year}{2014}), \eprint{1310.0492}.

\bibitem[{\citenamefont{{Shi} and {Krolik}}(2014)}]{Shi14}
\bibinfo{author}{\bibfnamefont{J.-M.} \bibnamefont{{Shi}}} \bibnamefont{and}
  \bibinfo{author}{\bibfnamefont{J.~H.} \bibnamefont{{Krolik}}},
  \bibinfo{journal}{in preparation}  (\bibinfo{year}{2014}).

\bibitem[{\citenamefont{{Nixon} et~al.}(2011)\citenamefont{{Nixon}, {Cossins},
  {King}, and {Pringle}}}]{Nixon11}
\bibinfo{author}{\bibfnamefont{C.~J.} \bibnamefont{{Nixon}}},
  \bibinfo{author}{\bibfnamefont{P.~J.} \bibnamefont{{Cossins}}},
  \bibinfo{author}{\bibfnamefont{A.~R.} \bibnamefont{{King}}},
  \bibnamefont{and} \bibinfo{author}{\bibfnamefont{J.~E.}
  \bibnamefont{{Pringle}}}, \bibinfo{journal}{\mnras}
  \textbf{\bibinfo{volume}{412}}, \bibinfo{pages}{1591} (\bibinfo{year}{2011}),
  \eprint{1011.1914}.

\bibitem[{\citenamefont{{Bankert} et~al.}(2014)\citenamefont{{Bankert},
  {Krolik}, and {Shi}}}]{Bankert14}
\bibinfo{author}{\bibfnamefont{J.}~\bibnamefont{{Bankert}}},
  \bibinfo{author}{\bibfnamefont{J.~H.} \bibnamefont{{Krolik}}},
  \bibnamefont{and} \bibinfo{author}{\bibfnamefont{J.~M.} \bibnamefont{{Shi}}},
  \bibinfo{journal}{in preparation}  (\bibinfo{year}{2014}).

\bibitem[{\citenamefont{{Bode} et~al.}(2010)\citenamefont{{Bode}, {Haas},
  {Bogdanovi{\'c}}, {Laguna}, and {Shoemaker}}}]{Bode10}
\bibinfo{author}{\bibfnamefont{T.}~\bibnamefont{{Bode}}},
  \bibinfo{author}{\bibfnamefont{R.}~\bibnamefont{{Haas}}},
  \bibinfo{author}{\bibfnamefont{T.}~\bibnamefont{{Bogdanovi{\'c}}}},
  \bibinfo{author}{\bibfnamefont{P.}~\bibnamefont{{Laguna}}}, \bibnamefont{and}
  \bibinfo{author}{\bibfnamefont{D.}~\bibnamefont{{Shoemaker}}},
  \bibinfo{journal}{\apj} \textbf{\bibinfo{volume}{715}}, \bibinfo{pages}{1117}
  (\bibinfo{year}{2010}), \eprint{0912.0087}.

\bibitem[{\citenamefont{{Bode} et~al.}(2012)\citenamefont{{Bode},
  {Bogdanovi{\'c}}, {Haas}, {Healy}, {Laguna}, and {Shoemaker}}}]{Bode12}
\bibinfo{author}{\bibfnamefont{T.}~\bibnamefont{{Bode}}},
  \bibinfo{author}{\bibfnamefont{T.}~\bibnamefont{{Bogdanovi{\'c}}}},
  \bibinfo{author}{\bibfnamefont{R.}~\bibnamefont{{Haas}}},
  \bibinfo{author}{\bibfnamefont{J.}~\bibnamefont{{Healy}}},
  \bibinfo{author}{\bibfnamefont{P.}~\bibnamefont{{Laguna}}}, \bibnamefont{and}
  \bibinfo{author}{\bibfnamefont{D.}~\bibnamefont{{Shoemaker}}},
  \bibinfo{journal}{\apj} \textbf{\bibinfo{volume}{744}}, \bibinfo{eid}{45}
  (\bibinfo{year}{2012}), \eprint{1101.4684}.

\bibitem[{\citenamefont{{Palenzuela} et~al.}(2010)\citenamefont{{Palenzuela},
  {Lehner}, and {Yoshida}}}]{Pal10b}
\bibinfo{author}{\bibfnamefont{C.}~\bibnamefont{{Palenzuela}}},
  \bibinfo{author}{\bibfnamefont{L.}~\bibnamefont{{Lehner}}}, \bibnamefont{and}
  \bibinfo{author}{\bibfnamefont{S.}~\bibnamefont{{Yoshida}}},
  \bibinfo{journal}{\prd} \textbf{\bibinfo{volume}{81}}, \bibinfo{eid}{084007}
  (\bibinfo{year}{2010}), \eprint{0911.3889}.

\bibitem[{\citenamefont{{Farris} et~al.}(2010)\citenamefont{{Farris}, {Liu},
  and {Shapiro}}}]{Farris10}
\bibinfo{author}{\bibfnamefont{B.~D.} \bibnamefont{{Farris}}},
  \bibinfo{author}{\bibfnamefont{Y.~T.} \bibnamefont{{Liu}}}, \bibnamefont{and}
  \bibinfo{author}{\bibfnamefont{S.~L.} \bibnamefont{{Shapiro}}},
  \bibinfo{journal}{\prd} \textbf{\bibinfo{volume}{81}}, \bibinfo{eid}{084008}
  (\bibinfo{year}{2010}), \eprint{0912.2096}.

\bibitem[{\citenamefont{{Farris} et~al.}(2012)\citenamefont{{Farris}, {Gold},
  {Paschalidis}, {Etienne}, and {Shapiro}}}]{Farris12}
\bibinfo{author}{\bibfnamefont{B.~D.} \bibnamefont{{Farris}}},
  \bibinfo{author}{\bibfnamefont{R.}~\bibnamefont{{Gold}}},
  \bibinfo{author}{\bibfnamefont{V.}~\bibnamefont{{Paschalidis}}},
  \bibinfo{author}{\bibfnamefont{Z.~B.} \bibnamefont{{Etienne}}},
  \bibnamefont{and} \bibinfo{author}{\bibfnamefont{S.~L.}
  \bibnamefont{{Shapiro}}}, \bibinfo{journal}{Physical Review Letters}
  \textbf{\bibinfo{volume}{109}}, \bibinfo{eid}{221102} (\bibinfo{year}{2012}),
  \eprint{1207.3354}.

\bibitem[{\citenamefont{{Giacomazzo} et~al.}(2012)\citenamefont{{Giacomazzo},
  {Baker}, {Miller}, {Reynolds}, and {van Meter}}}]{Giacomazzo12}
\bibinfo{author}{\bibfnamefont{B.}~\bibnamefont{{Giacomazzo}}},
  \bibinfo{author}{\bibfnamefont{J.~G.} \bibnamefont{{Baker}}},
  \bibinfo{author}{\bibfnamefont{M.~C.} \bibnamefont{{Miller}}},
  \bibinfo{author}{\bibfnamefont{C.~S.} \bibnamefont{{Reynolds}}},
  \bibnamefont{and} \bibinfo{author}{\bibfnamefont{J.~R.} \bibnamefont{{van
  Meter}}}, \bibinfo{journal}{ArXiv e-prints}  (\bibinfo{year}{2012}),
  \eprint{1203.6108}.

\bibitem[{\citenamefont{Gold et~al.}(2013)\citenamefont{Gold, Paschalidis,
  Etienne, Shapiro, and Pfeiffer}}]{Gold:2013zma}
\bibinfo{author}{\bibfnamefont{R.}~\bibnamefont{Gold}},
  \bibinfo{author}{\bibfnamefont{V.}~\bibnamefont{Paschalidis}},
  \bibinfo{author}{\bibfnamefont{Z.~B.} \bibnamefont{Etienne}},
  \bibinfo{author}{\bibfnamefont{S.~L.} \bibnamefont{Shapiro}},
  \bibnamefont{and} \bibinfo{author}{\bibfnamefont{H.~P.}
  \bibnamefont{Pfeiffer}} (\bibinfo{year}{2013}), \eprint{1312.0600}.

\bibitem[{\citenamefont{Blanchet et~al.}(1998)\citenamefont{Blanchet, Faye, and
  Ponsot}}]{Blanchet:1998vx}
\bibinfo{author}{\bibfnamefont{L.}~\bibnamefont{Blanchet}},
  \bibinfo{author}{\bibfnamefont{G.}~\bibnamefont{Faye}}, \bibnamefont{and}
  \bibinfo{author}{\bibfnamefont{B.}~\bibnamefont{Ponsot}},
  \bibinfo{journal}{Phys. Rev.} \textbf{\bibinfo{volume}{D58}},
  \bibinfo{pages}{124002} (\bibinfo{year}{1998}), \eprint{gr-qc/9804079}.

\bibitem[{\citenamefont{{Gammie} et~al.}(2003)\citenamefont{{Gammie},
  {McKinney}, and {T{\'o}th}}}]{GMT03}
\bibinfo{author}{\bibfnamefont{C.~F.} \bibnamefont{{Gammie}}},
  \bibinfo{author}{\bibfnamefont{J.~C.} \bibnamefont{{McKinney}}},
  \bibnamefont{and}
  \bibinfo{author}{\bibfnamefont{G.}~\bibnamefont{{T{\'o}th}}},
  \bibinfo{journal}{\apj} \textbf{\bibinfo{volume}{589}}, \bibinfo{pages}{444}
  (\bibinfo{year}{2003}), \eprint{astro-ph/0301509}.

\bibitem[{\citenamefont{Noble et~al.}(2009)\citenamefont{Noble, Krolik, and
  Hawley}}]{Noble:2008tm}
\bibinfo{author}{\bibfnamefont{S.~C.} \bibnamefont{Noble}},
  \bibinfo{author}{\bibfnamefont{J.~H.} \bibnamefont{Krolik}},
  \bibnamefont{and} \bibinfo{author}{\bibfnamefont{J.~F.}
  \bibnamefont{Hawley}}, \bibinfo{journal}{Astrophys. J.}
  \textbf{\bibinfo{volume}{692}}, \bibinfo{pages}{411} (\bibinfo{year}{2009}),
  \eprint{0808.3140}.

\bibitem[{\citenamefont{Will}(2011)}]{Will:2011nz}
\bibinfo{author}{\bibfnamefont{C.~M.} \bibnamefont{Will}},
  \bibinfo{journal}{Proc. Nat. Acad. Sci.} \textbf{\bibinfo{volume}{108}},
  \bibinfo{pages}{5938} (\bibinfo{year}{2011}), \eprint{1102.5192}.

\bibitem[{\citenamefont{Mundim et~al.}(2014)\citenamefont{Mundim, Nakano,
  Yunes, Campanelli, Noble et~al.}}]{Mundim:2013vca}
\bibinfo{author}{\bibfnamefont{B.~C.} \bibnamefont{Mundim}},
  \bibinfo{author}{\bibfnamefont{H.}~\bibnamefont{Nakano}},
  \bibinfo{author}{\bibfnamefont{N.}~\bibnamefont{Yunes}},
  \bibinfo{author}{\bibfnamefont{M.}~\bibnamefont{Campanelli}},
  \bibinfo{author}{\bibfnamefont{S.~C.} \bibnamefont{Noble}},
  \bibnamefont{et~al.}, \bibinfo{journal}{Phys. Rev.}
  \textbf{\bibinfo{volume}{D89}}, \bibinfo{pages}{084008}
  (\bibinfo{year}{2014}), \eprint{1312.6731}.

\bibitem[{\citenamefont{Zilhao and Noble}(2014)}]{Zilhao:2013dta}
\bibinfo{author}{\bibfnamefont{M.}~\bibnamefont{Zilhao}} \bibnamefont{and}
  \bibinfo{author}{\bibfnamefont{S.~C.} \bibnamefont{Noble}},
  \bibinfo{journal}{Class. Quant. Grav.} \textbf{\bibinfo{volume}{31}},
  \bibinfo{pages}{065013} (\bibinfo{year}{2014}), \eprint{1309.2960}.

\bibitem[{\citenamefont{Misner et~al.}(1973)\citenamefont{Misner, Thorne, and
  Wheeler}}]{Misner73}
\bibinfo{author}{\bibfnamefont{C.~W.} \bibnamefont{Misner}},
  \bibinfo{author}{\bibfnamefont{K.~S.} \bibnamefont{Thorne}},
  \bibnamefont{and} \bibinfo{author}{\bibfnamefont{J.~A.}
  \bibnamefont{Wheeler}}, \emph{\bibinfo{title}{Gravitation}}
  (\bibinfo{publisher}{W. H. Freeman}, \bibinfo{address}{San Francisco},
  \bibinfo{year}{1973}).

\bibitem[{\citenamefont{{T{\'o}th}}(2000)}]{2000JCoPh.161..605T}
\bibinfo{author}{\bibfnamefont{G.}~\bibnamefont{{T{\'o}th}}},
  \bibinfo{journal}{Journal of Computational Physics}
  \textbf{\bibinfo{volume}{161}}, \bibinfo{pages}{605} (\bibinfo{year}{2000}).

\bibitem[{\citenamefont{{Noble} et~al.}(2006)\citenamefont{{Noble}, {Gammie},
  {McKinney}, and {Del Zanna}}}]{Noble06}
\bibinfo{author}{\bibfnamefont{S.~C.} \bibnamefont{{Noble}}},
  \bibinfo{author}{\bibfnamefont{C.~F.} \bibnamefont{{Gammie}}},
  \bibinfo{author}{\bibfnamefont{J.~C.} \bibnamefont{{McKinney}}},
  \bibnamefont{and} \bibinfo{author}{\bibfnamefont{L.}~\bibnamefont{{Del
  Zanna}}}, \bibinfo{journal}{\apj} \textbf{\bibinfo{volume}{641}},
  \bibinfo{pages}{626} (\bibinfo{year}{2006}), \eprint{arXiv:astro-ph/0512420}.

\bibitem[{\citenamefont{Farris et~al.}(2011)\citenamefont{Farris, Liu, and
  Shapiro}}]{Farris:2011vx}
\bibinfo{author}{\bibfnamefont{B.~D.} \bibnamefont{Farris}},
  \bibinfo{author}{\bibfnamefont{Y.~T.} \bibnamefont{Liu}}, \bibnamefont{and}
  \bibinfo{author}{\bibfnamefont{S.~L.} \bibnamefont{Shapiro}},
  \bibinfo{journal}{Phys. Rev.} \textbf{\bibinfo{volume}{D84}},
  \bibinfo{pages}{024024} (\bibinfo{year}{2011}), \eprint{1105.2821}.

\bibitem[{\citenamefont{Noble et~al.}(2010)\citenamefont{Noble, Krolik, and
  Hawley}}]{Noble:2010mm}
\bibinfo{author}{\bibfnamefont{S.~C.} \bibnamefont{Noble}},
  \bibinfo{author}{\bibfnamefont{J.~H.} \bibnamefont{Krolik}},
  \bibnamefont{and} \bibinfo{author}{\bibfnamefont{J.~F.}
  \bibnamefont{Hawley}}, \bibinfo{journal}{Astrophys. J.}
  \textbf{\bibinfo{volume}{711}}, \bibinfo{pages}{959} (\bibinfo{year}{2010}),
  \eprint{1001.4809}.

\bibitem[{\citenamefont{{Hawley} et~al.}(2011)\citenamefont{{Hawley}, {Guan},
  and {Krolik}}}]{hgk11}
\bibinfo{author}{\bibfnamefont{J.~F.} \bibnamefont{{Hawley}}},
  \bibinfo{author}{\bibfnamefont{X.}~\bibnamefont{{Guan}}}, \bibnamefont{and}
  \bibinfo{author}{\bibfnamefont{J.~H.} \bibnamefont{{Krolik}}},
  \bibinfo{journal}{\apj} \textbf{\bibinfo{volume}{738}}, \bibinfo{eid}{84}
  (\bibinfo{year}{2011}), \eprint{1103.5987}.

\bibitem[{\citenamefont{{Sorathia} et~al.}(2012)\citenamefont{{Sorathia},
  {Reynolds}, {Stone}, and {Beckwith}}}]{2012ApJ...749..189S}
\bibinfo{author}{\bibfnamefont{K.~A.} \bibnamefont{{Sorathia}}},
  \bibinfo{author}{\bibfnamefont{C.~S.} \bibnamefont{{Reynolds}}},
  \bibinfo{author}{\bibfnamefont{J.~M.} \bibnamefont{{Stone}}},
  \bibnamefont{and}
  \bibinfo{author}{\bibfnamefont{K.}~\bibnamefont{{Beckwith}}},
  \bibinfo{journal}{\apj} \textbf{\bibinfo{volume}{749}}, \bibinfo{eid}{189}
  (\bibinfo{year}{2012}), \eprint{1106.4019}.

\end{thebibliography}

\end{document}